\documentclass[reprint,superscriptaddress,amsmath,amssymb,aps,pra]{revtex4-2}

\usepackage{amsfonts}
\usepackage{amsmath}
\usepackage{amsthm}
\usepackage{mathrsfs}
\usepackage{amscd}
\usepackage{amssymb}
\usepackage{subfigure}
\usepackage{amsxtra}
\usepackage{dcolumn}
\usepackage{bm}           
\usepackage{bbm}
\usepackage{graphicx}
\usepackage{epstopdf}
\usepackage{color}
\usepackage[dvipsnames]{xcolor}
\usepackage[colorlinks,linkcolor=Blue,citecolor=Green]{hyperref}

\newcommand{\la}{\langle}
\newcommand{\ra}{\rangle}

\newcommand{\Op}[1]{\hat{#1}}

\newcommand{\osigma}{\Op{\sigma}}

\newcommand{\oH}{\Op{H}}
\newcommand{\oA}{\Op{A}}
\newcommand{\oB}{\Op{B}}

\newcommand{\oO}{\Op{O}}

\newcommand{\oa}{\Op{a}}

\newcommand{\tr}{\ensuremath{{\rm tr}}}

\newcommand{\abs}[1]{\left|#1\right|}

\usepackage{xcolor}

\begin{document}

\preprint{APS/123-QED}

\title{Non-Markovian dynamics of the giant atom beyond the rotating-wave approximation}
\author{Mei Yu}
\email{mei.yu@tu-dresden.de}
\affiliation{Institut für Theoretische Physik, Technische Universität Dresden, D-01062 Dresden, Germany}

\author{Walter T. Strunz}
\affiliation{Institut für Theoretische Physik, Technische Universität Dresden, D-01062 Dresden, Germany}

\author{Stefan Nimmrichter}
\affiliation{Naturwissenschaftlich-Technische Fakult{\"a}t, Universit{\"a}t Siegen, Siegen 57068, Germany}

\begin{abstract}
We study the non-Markovian dynamics of a giant artificial atom coupled to a one-dimensional acoustic waveguide beyond the rotating-wave and weak-coupling approximations. By combining an optimized ESPRIT-based decomposition of the bath correlation function with the hierarchical equations of motion (HEOM), we achieve numerically exact simulations in regimes with long memory times, finite temperature, and strong system–bath coupling. Benchmarking against analytical results reveals the breakdown of perturbative non-Markovian approaches such as Redfield theory even at weak coupling in the presence of delay-induced memory. We further show that non-Markovian features, including excitation revivals, remain robust at finite temperature and can be enhanced by increasing the system-bath coupling strength. Our approach provides a versatile framework for studying non-Markovian quantum dynamics in structured environments relevant to giant-atom platforms.
\end{abstract}
\maketitle

\section{Introduction}
Recent advances in circuit quantum electrodynamics have enabled the realization of giant artificial atoms---superconducting qubits coupled to acoustic or photonic waveguides at multiple spatially separated points \cite{Du2022, Vadiraj2021, Kannan2020, Manenti2017, Andersson2019}. 
In contrast to conventional dipole-like light–matter interactions, the spatially extended coupling gives rise to interference effects that enable frequency-dependent decay rates and Lamb shifts \cite{Frisk2014}, the formation of bound states in the continuum \cite{Guo2020_1, Guo2020_2, Zhao2020,Noachtar2022}, and nonreciprocal single-photon scattering \cite{Zhou2023}. Moreover, giant atoms can be engineered to support decoherence-free interactions \cite{Kockum2018, Du2023} and enable the generation of entanglement \cite{Yin2022, Yin2023, Yu2021, Santos2023} between them.
Beyond these effects, giant atoms provide a promising platform for open-system quantum simulation \cite{Chen2025} and for realizing giant superatoms with protected entanglement manipulation and routing \cite{Du2025}.

A key feature of giant atoms is the emergence of time-delayed coherent feedback \cite{Grimsmo2015, Pichler2016}. Due to the finite propagation speed of excitations in the waveguide, emission at one coupling point can be reabsorbed at another after a finite delay that may exceed the atomic decay time. This leads to pronounced non-Markovian dynamics, where the system evolution depends on its past history \cite{Caruso2014, Breuer2016, deVega2017}. Such delay-induced memory effects have been experimentally observed, for instance through non-exponential relaxation dynamics \cite{Andersson2019}, and can be engineered by tuning the distance between coupling points or the properties of the waveguide \cite{Kannan2020}. Moreover, the memory effect is phase coherent and thus genuinely nonclassical \cite{Yu2025}, which could aid the development of quantum memory systems that preserve coherence over longer timescales \cite{Wang2017} and enable cycles of coherent energy exchange, for use in quantum thermal machines \cite{Binder2019, Newman2017} and entanglement harvesting \cite{Valentin1991, Juan2009, Sabin2010}.

\begin{figure}[t!]
    \centering
        \includegraphics[width=0.8\linewidth]{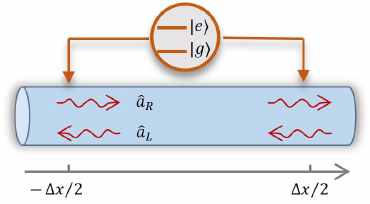}
    \caption{Two-level giant atom coupled to the left- and right- propagating modes in a 1D acoustic waveguide via multiple contacts, with a separation of $\Delta x$.}
    \label{model_sketch}
\end{figure}

Despite these advances, accurately describing the dynamics of giant atoms remains a major theoretical challenge, as strong memory effects arising from the time-delayed feedback lead to the breakdown of conventional perturbative Markovian methods. Such delay-induced coherent feedback is also known in waveguide quantum electrodynamics, including paradigmatic settings such as an atom in front of a mirror, where finite propagation time gives rise to non-Markovian dynamics and invalidates Markovian (Lindblad-type) master equations \cite{ Grimsmo2015,Pichler2016}. Giant atoms provide a particularly rich realization of this physics, as multiple coupling points generate a highly structured bath correlation function with multiple time-delayed contributions.
While analytical solutions exist in the weak-coupling and zero-temperature limit within the rotating-wave approximation (RWA) \cite{Guo2017}, it is challenging to extend beyond this regime and include multi-phonon processes that arise, for example, at strong coupling or at finite temperature.

Numerically exact approaches, such as matrix-product-state (MPS) simulations \cite{Grimsmo2015,Magnifico2025}, the time evolving density operator with orthogonal polynomials algorithm (TEDOPA) \cite{Chin2010}, and the time-evolving matrix product operator (TEMPO) method \cite{Strathearn2018,Link2024}, or hierarchical approaches in terms of density operators (HEOM), or stochastic pure states (HOPS) \cite{Suess2014}  may offer potential routes toward tackling the fully non-perturbative dynamics. In particular, Ref.\ \cite{Noachtar2022} demonstrated how to access the zero-temperature giant-atom dynamics beyond the RWA with the help of a combined chain-mapping and MPS technique. However, the explicit representation of bath degrees of freedom in terms of a one-dimensional chain of harmonic modes leads to rapidly growing entanglement, making simulations increasingly demanding for long-time dynamics, finite-temperature regimes, and environments with long memory times. A unified and efficient treatment of these regimes thus remains challenging, and even more so at finite temperature. In this context, a further important open question is whether non-Markovian perturbative approaches, such as the Redfield master equation, stay reliable in the presence of strongly structured, delay-induced bath correlations, or whether fully non-perturbative methods are required.

In this work, we address these challenges using the hierarchical equations of motion (HEOM) formalism \cite{Tanimura1989, Tanimura2020, Lambert2023}, combined with a tailored and optimized exponential decomposition scheme for the bath correlation function (BCF) that we develop based on the ESPRIT algorithm \cite{Roy1989, Takahashi2024}. The HEOM approach captures non-Markovian dynamics through a hierarchy of auxiliary density operators, and our ESPRIT-based scheme enables an accurate and efficient representation of the highly structured, long-memory correlation functions characteristic of giant-atom systems.
It allows us to access a regime combining long-time dynamics, long bath memory, finite temperature, and strong system–bath coupling that is challenging for existing methods. 

We first benchmark our approach against the known analytical result for zero temperature and demonstrate the failure of the perturbative Redfield master equation to capture delay-induced memory effects even at weak coupling strengths. Furthermore, we show that non-Markovian features, such as population revivals, persist at finite temperature.
 This robustness may be relevant for applications in quantum thermodynamics, for instance in the design of thermal machines. In addition, we find that increasing the coupling strength can reinforce the memory,  highlighting the interplay between strong coupling, time-delayed coherent feedback, and thermal fluctuations in structured environments at finite temperatures.

The article is organized as follows: Section~\ref{sec:physical_model} introduces the physical model of a two-level giant atom coupled to a one-dimensional surface acoustic waveguide and discusses the challenges in obtaining an exact description of its reduced dynamics. Section~\ref{sec:HEOM_method} presents the HEOM approach used for numerical simulations.  
Section~\ref{sec:BCF_exp_fit} introduces the optimized ESPRIT fitting scheme and demonstrates its high accuracy for the considered spectral density. Section~\ref{sec:benchmark} benchmarks the HEOM results against analytical and perturbative solutions, while Sec.~\ref{sec:multiphonon_regime} extends the method to finite temperatures and strong coupling. Conclusions are given in Sec.~\ref{sec:conclusion}.

\section{Giant atom model} \label{sec:physical_model}
The model under consideration is based on the physical setup of Ref.~\cite{Andersson2019} in which the non-Markovian excitation revivals of a giant atom were clearly observed for the first time. A superconducting transmon qubit with resonance frequency $\omega_0$ couples via two contact points of distance $\Delta x$ to a transmission line of surface acoustic waves (SAWs) of velocity $v_g$ on a piezoelectric substrate, as sketched in Fig.~\ref{model_sketch}. 
The acoustic wavelength at the qubit resonance frequency is several orders of magnitude smaller than $\Delta x$. Introducing the characteristic time delay, $\tau = \Delta x/v_g$, we have $\omega_0 \tau \gg 1$. The spectral coupling rate between the qubit and a running-wave SAW mode of frequency $\omega$ modulates with the phase difference between the two contact points, ranging from a maximum $\gamma(\omega)$ whenever $\omega\tau = 0 \mod{2\pi}$ to zero when $\omega\tau = \pi \mod{2\pi}$. The corresponding Hamiltonian reads as \cite{Guo2017},
\begin{eqnarray}
\oH &=&  \hbar \omega_0 |e\rangle \langle e| + \hbar \sum_{j=R,L} \int_0^{\infty} d\omega \,\omega \hat{a}_{j\omega}^{\dagger} \hat{a}_{j\omega} \\
&+&\hbar \sum_{j=R,L} \int_0^{\infty} d\omega \sqrt{\frac{\gamma(\omega)}{\pi}} \cos{\left(\frac{\omega \tau}{2}\right)} \osigma_x  \left( \hat{a}_{j\omega}^{\dagger} + \oa_{j\omega} \right). \nonumber 
\label{total_hamiltonian}
\end{eqnarray} 
Here, the operator $\osigma_x = \osigma_-+\osigma_+$ governs the interaction with the field, with $\osigma_- = |g\rangle \langle e|$ and $\osigma_+ = |e\rangle \langle g |$ the lowering and raising operators of the giant atom. The field operators $\hat{a}_{j\omega}$ describe a continuum of right- ($j=R$) and left-propagating ($j=L$) SAW phonon modes, obeying the commutation relation $[\hat{a}_{j\omega}, \hat{a}_{j'\omega'}^{\dagger}] = \delta_{jj'} \delta(\omega-\omega')$.
In the interaction picture, the Hamiltonian reduces to $\oH_I (t) = \hbar \osigma_x (t) \oB(t)$, with $\osigma_x(t) = \osigma_- e^{-i\omega_0 t} + h.c.$ and 
\begin{equation}\label{eq:B}
    \oB (t) = \sum_{j=R,L} \int_0^{\infty} d\omega \sqrt{\frac{\gamma(\omega)}{\pi}} \cos{\left(\frac{\omega \tau}{2}\right)} \hat{a}_{j\omega}^{\dagger} e^{i\omega t} + h.c. 
\end{equation}
The memory effects of the field on the atom are fully characterized by BCF, given by
\begin{eqnarray}\label{eq:BCF_gene}
    C(t) &=& \la \oB(t) \oB(0)\ra =
\frac{1}{\pi} \int_{-\infty}^{\infty} d\omega J(\beta, \omega) e^{-i\omega t}, \\
&& J(\beta, \omega) = \frac{2\gamma(\abs{\omega})}{\abs{1-e^{-\beta \omega}}}\cos^2{\left(\frac{\omega \tau}{2}\right)}. \nonumber
\end{eqnarray}
Here, the field is initially assumed to be in a thermal state with inverse temperature $\beta =\hbar/k_B T$.  The effective bath spectral density function $J(\beta, \omega)$ exhibits rapid oscillations on top of the slowly varying $\gamma(\omega)$ due to interference between the two contact points. 
Hence, since the BCF is the Fourier transform of the effective spectral density, it features pronounced peaks at $t=0$ and $t=\pm \tau$. 
We adopt an Ohmic spectral density of the form $\gamma(\omega) = \eta \omega e^{-\omega/\omega_c}$ with a dimensionless strength parameter $\eta$ and a cutoff frequency $\omega_c$. This form follows directly from the microscopic qubit-SAW coupling in one dimension \cite{Guo2020}: the interaction strength scales as $g(k) \propto \sqrt{|k|}$ for each $k$-mode and together with the linear dispersion relation  $\omega = v_g|k|$ of SAWs, leads to a spectral density that is linear in frequency, $\gamma(\omega) \propto \omega$. The exponential cutoff models the finite bandwidth of the environment by suppressing contributions from high-frequency modes, where the assumptions of linear dispersion and efficient coupling break down. It provides a smooth regularization of the spectral density while preserving the low-frequency Ohmic behavior without introducing an artificial sharp boundary \cite{Leggett1987}. The cutoff frequency $\omega_c$ defines the energy scale of this suppression and controls the effective temporal width of the BCF: larger $\omega_c$ leads to faster decay of the BCF, whereas smaller $\omega_c$ results in longer bath memory times and stronger non-Markovian effects; see App.~\ref{app:dependence_cutoff_freqs.} for details. While other smooth cutoffs, such as Gaussian functions, could also be employed, we emphasize that the qualitative behavior reported in this work is not sensitive to the specific form of the cutoff function.
The effective bath spectral density and the BCF are plotted in Fig.~\ref{SD_BCF_plot} at zero temperature. 
As a result, in this setting, the field and the atom remain strongly correlated for an extended period, particularly when the time delay $\tau$ is comparable to or longer than the atom's relaxation time. In the Markovian regime of zero delay and weak coupling, $\tau = 0$ and $\gamma(\omega_0) \ll \omega_0$, the atom relaxes at a rate of $\gamma(\omega_0)$. Significant memory effects beyond the atomic lifetime are therefore expected to occur for $\gamma(\omega_0)\tau \gtrsim 1$, making it difficult in general to accurately describe the dynamics of the reduced atomic state.

\section{HEOM-based simulated giant atom dynamics}\label{sec:HEOM_method}
To simulate the dynamics of the giant atom, we employ the HEOM method, a non-perturbative approach designed to handle non-Markovian dynamics in open systems, particularly in the strong-coupling regime or when pronounced memory effects are present. 
The HEOM formalism relies on the assumption that the real and imaginary parts of the BCF can be approximated as a sum of exponentials,
\begin{eqnarray}\label{bcf_exp_representation}
  C_{\mathrm{R}}(t) &\approx& \sum_{k=1} ^{N_{\mathrm{R}}} c_k^{\mathrm{R}} e^{-\gamma_k^{\mathrm{R}}t}, ~
  C_{\mathrm{I}}(t) \approx \sum_{k=1} ^{N_{\mathrm{I}}} c_k^{\mathrm{I}} e^{-\gamma_k^{\mathrm{I}}t},
\end{eqnarray}
where $c_k^{j}$ and $\gamma_k^{j}$ can be real or complex. 
Under this decomposition, the system’s exact dynamics can be systematically derived using the influence functional approach \cite{Tanimura1989, Tanimura2020, Lambert2023}. Given an initial product state of system and phonon bath, $\rho(0)=\rho_S(0)\otimes \rho_B(0)$, one arrives at a hierarchy of coupled equations,
\begin{eqnarray} \label{eq:heom_eqs} 
    \frac{d}{dt} \rho_S^n(t) &=& \left( -i\omega_0 |e\rangle \langle e|^{\times} - \sum_{j=\mathrm{R}, \mathrm{I}}  \sum^{N_j}_{k=1} n_{jk}\gamma_k^j\right) \rho_S^{n}(t)  \nonumber \\
    &-& i\sum_{k=1}^{N_\mathrm{R}} c_k^{\mathrm{R}}n_{\mathrm{R}k} \osigma_x^{\times}\rho_S^{n^-_{\mathrm{R}k}}
    + \sum_{k=1}^{N_\mathrm{I}} c_k^{\mathrm{I}}n_{\mathrm{I}k} \osigma_x^{\circ}\rho_S^{n^-_{\mathrm{I}k}}(t) \nonumber \\
    &-& i\sum_{j=\mathrm{R}, \mathrm{I}}  \sum^{N_j}_{k=1} \osigma_x^{\times} \rho_S^{n^+_{jk}} (t).
\end{eqnarray}
Here, the superoperators $O^{\times}$ and $O^{\circ}$ are defined as $O^{\times} \oA = \big[\oO, \oA \big]$ and $O^{\circ} \oA = \big\{\oO, \oA \big\}$, representing the commutator and anti-commutator actions, respectively.
The multi-index $n$ is given by $n= \left(n_{\mathrm{R}1}, n_{\mathrm{R}2}, \ldots, n_{\mathrm{R}N_{\mathrm{R}}}, n_{\mathrm{I}1}, n_{\mathrm{I}2},\ldots, n_{\mathrm{I}N_{\mathrm{I}}} \right)$, where each component $n_{jk}$ is an integer within the range $\left[0, N_c\right]$, with $N_c$ representing the truncation order for the numerical convergence. The density operator corresponding to $n=(0,0,\ldots 0)$ represents the reduced system state, while density operators with non-zero indices, referred to as auxiliary density operators (ADOs), encode the memory effect of the environment.  The notation $\rho_S^{n^{\pm}_{jk}}(t)$ indicates an ADO with a specific index either increased or decreased by one. 

Although the hierarchy formally contains an infinite number of equations, practical computations require truncation at a finite hierarchy depth $N_c$, beyond which higher-order terms have negligible contributions. The resulting finite set of coupled differential equations can be efficiently solved using standard numerical methods, as conveniently implemented in the HEOM module of the QuTiP framework \cite{Lambert2023}. 
Truncating the exponential decomposition of the BCFs \eqref{bcf_exp_representation} can introduce numerical artifacts, potentially leading to unphysical results and reducing the accuracy of the simulation. At the same time, to ensure computational efficiency, it is crucial to represent the BCF with as few exponential terms as possible while maintaining sufficient accuracy. Striking a balance between precision and computational cost is essential for the effective application of the HEOM method. Our simulations typically required several hundreds of exponential terms to converge, with a HEOM depth of $N_c=2$. 

\section{Exponential BCF decomposition}\label{sec:BCF_exp_fit}
 Various numerical techniques have been developed to achieve an exponential decomposition of BCFs for use in HEOM, broadly categorized into frequency-domain and time-domain approaches. In the frequency domain, methods such as Matsubara frequency decomposition  \cite{Tanimura1990,Ishizaki2005, Ding2012} and Padé spectral decomposition \cite{Hu2010, Hu2011} are commonly applied to simple spectral densities. However, these methods often require numerous terms for convergence, particularly at low temperatures, where the Fano spectrum decomposition \cite{Cui2019, Zhang2020} and Free-pole decomposition \cite{Xu2022} are often more effective. Time-domain methods, including least squares fitting \cite{Vanhuffel1994, Aushev2014, Hartmann2019},  and Prony's method \cite{Chen2022, Schaubert1979}, excel in localized fitting and are well suited for short-to-medium time behavior. However, they may encounter difficulties when applied to highly structured or multi-peaked BCFs.

\begin{figure}[t!]
    \centering
        \includegraphics[width=1\linewidth]{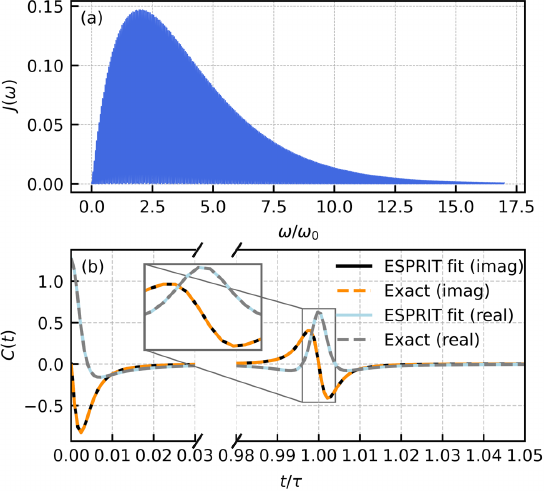}
    \caption{(a) Effective bath spectral density of the SAW field at zero temperature; (b) The corresponding bath correlation function as a function of time. Here and in the subsequent plots, we consider an Ohmic spectral density of the SAW environment, $\gamma(\omega) = \eta \omega e^{-\omega/\omega_c}$, with the coupling strength and cutoff frequency parameters set as $\eta = 1 \times 10^{-2}$, $\omega_c/\omega_0 = 2$ and $\omega_0\tau /2\pi = 20$.  The inset provides a zoomed-in view of the second peaks and their fits.}
    \label{SD_BCF_plot}
\end{figure}

Given the Ohmic spectral density $\gamma(\omega)$ we consider for both contact points, the BCF \eqref{eq:BCF_gene} takes the analytical form,
\begin{eqnarray} \label{eq:bcf_GA_OhmicSD}
C(t) &=& \frac{\eta}{2\pi \beta^2} \Biggl\{2\Psi\bigg[1,\frac{1+i t \omega_c}{\beta \omega_c}\bigg] + 2\Psi \bigg[1,\frac{r}{\beta\omega_c}\bigg] \nonumber \\
&+& \Psi \bigg[1,\frac{1+i\left(t-\tau\right) \omega_c}{\beta \omega_c}\bigg] + \Psi \bigg[1,\frac{1+i\left(t+\tau\right) \omega_c}{\beta \omega_c}\bigg] \nonumber \\
&+& \Psi \bigg[1,\frac{r+i\tau\omega_c}{\beta \omega_c}\bigg]
+ \Psi \bigg[1,\frac{r-i\tau \omega_c}{\beta \omega_c}\bigg] \Biggl\}.
\end{eqnarray}
with $r =1-i\tau\omega_c +\beta \omega_c$. 
Here, $\Psi[1,x]$ denotes the first derivative of the digamma function \cite{Abramowitz1965}.  
As illustrated in Fig.~\ref{SD_BCF_plot}(b), the BCF \eqref{eq:bcf_GA_OhmicSD} exhibits sharp peaks at distinct times (integer multiples of the delay time $\tau$), with a flat region between them. This structure poses significant challenges for the aforementioned decomposition methods, particularly as larger time delays accentuate the distinctness of the peaks, making an accurate representation more difficult.

In this work, we develop an optimized fitting scheme based on the ESPRIT algorithm~\cite{Roy1989, Takahashi2024}. The method is refined to efficiently represent both the real and imaginary parts of the BCF~\eqref{eq:bcf_GA_OhmicSD} with a sum of exponential functions as in Eq.~\eqref{bcf_exp_representation}.
The ESPRIT algorithm exploits the rotational invariance property of a Vandermonde matrix constructed from time-domain samples of the BCF. The procedure begins with sampling the BCF on a discrete-time grid, followed by the formation of a Hankel matrix from these samples.  
Singular value decomposition (SVD) is then applied to estimate the number of exponential components. In general, a direct ESPRIT fit tends to yield a redundant set of exponentials. To address this, we iteratively refine the fitted components by gradually reducing the number of terms while monitoring the relative fitting error. The iteration terminates once a closeness condition based on the $L^2$-norm is satisfied, $||C_{\mathrm{fit}}(t)-C(t)||_2/||C(t)||_2 < \varepsilon_r$, ensuring an accurate and compact exponential representation of the BCF.
Reducing the threshold $\varepsilon_r$ improves the accuracy of the fit, but requires a greater number of exponential functions to achieve this precision. Subsequently, the decay rates $\gamma_k^{\rm R}$ and $\gamma_k^{\rm I}$ are determined by solving an eigenvalue problem, while the corresponding amplitudes $c_k^{\rm R}$ and $c_k^{\rm I}$ are obtained via a least-squares fitting approach. By reconstructing the BCF as a sum of exponential functions, the optimized ESPRIT algorithm proves to be a robust and effective tool for time-domain analysis. This is demonstrated in Fig.~\ref{SD_BCF_plot}(b), where the real and imaginary components of the original and optimized ESPRIT-fitted BCF show excellent agreement. The plot emphasizes the non-zero regions, focusing on the short-time decay and the oscillatory behavior near $t/\tau \approx 1$, excluding the intermediate flat region for visual clarity. For this fitting, we set the relative error threshold at $\varepsilon_r = 10^{-3}$, balancing accuracy and computational efficiency.

While the optimized ESPRIT method effectively fits the BCF using a sum of exponentials, it often introduces a large number of terms, especially for a long time delay parameter $\tau$. This leads to a high-dimensional hierarchy in HEOM simulations, significantly increasing computational demands and making numerical integration more expensive. Thus, further optimizing the ESPRIT method to minimize the number of terms while preserving accuracy could be an important avenue for future research. Reducing the number of exponential components would enhance the computational efficiency of HEOM simulations, making numerical integration less demanding and more practical for larger systems or extended time frames.

Using the fitting results, we then simulate the dynamics of a two-level giant atom with the HEOM implementation in QuTiP.
To validate our numerical approach, we first perform a benchmark at zero temperature, where an analytical solution is available in the RWA regime. In addition, we compare our results with those obtained from the Redfield master equation, in order to assess the validity of non-Markovian perturbative approaches in the presence of delay-induced memory effects.

\section{Benchmark at zero temperature}\label{sec:benchmark}
We shall investigate the dynamics of a two-level giant atom at zero temperature using different methods, focusing on the weak coupling regime. In the absence of thermal excitations, spontaneous emission arises due to the atom’s interaction with vacuum fluctuations of the acoustic field. Under these conditions, the rotating wave approximation is justified and the analysis can be confined to the single-excitation subspace of the field. See App.~\ref{app:TLS_GA_exa_ME_zero_temp} for a derivation of the exact time evolution in this setting. Equivalently, the time evolution of the giant atom state can be expressed in terms of a time-local non-Markovian master equation,
\begin{eqnarray} \label{eq:exact_me_GA}
    \frac{d \rho_S}{dt} &=& -i \left(\omega_0 + h(t) \right) \left[\osigma_+ \osigma_-, \rho_S\right] + \gamma_{-} (t) \mathcal{D}\left[\osigma_-\right]\rho_S,
\end{eqnarray}
where $h(t) = -\mathrm{Im} [\dot{G}(t)/G(t)]$ is the Lamb shift, which renormalizes the system energy, and $\mathcal{D}[\osigma_-]\rho_S(t) = 2 \osigma_- \rho_S(t)\osigma_+ - \{\osigma_+ \osigma_-, \rho_S(t)\}$ is the dissipator describing the spontaneous emission of the giant atom with a time-dependent decay rate $\gamma_-(t) = -\mathrm{Re} [\dot{G}(t)/G(t)]$. The Green function $G(t)$ satisfies the equation
\begin{eqnarray}\label{G_diff_funct}
    \dot{G}(t) &=& -\int_0^t ds\, C(s)G(t-s)e^{i\omega_0 s},
\end{eqnarray}
with initial condition $G(0)=1$. This time-local master equation is exact in the sense that its generator contains the full Dyson series in the coupling strength, including all higher-order terms. This is crucial in regimes where the bath correlation time is sufficiently large that higher-order contributions cannot be neglected \cite{vanKampen2011}.
The non-Markovianity of the master equation is signified by transient negativities of $\gamma_-(t)$ that may occur for $\tau>0$ \cite{Hall2014, deVega2017}. 

\begin{figure}[t!]
    \centering
        \includegraphics[width=1\linewidth]{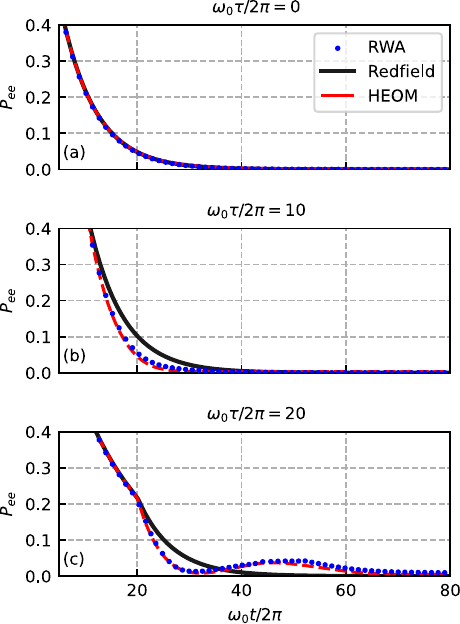}
    \caption{Excited-state probability of the giant atom as a function of time, obtained from the RWA analytical solution, the Redfield master equation, and from HEOM simulations at zero temperature. We compare three different time delays $\tau$: (a) zero time delay, which amounts to a single contact point; (b) time delay comparable to the atomic relaxation time; (c) time delay longer than the atomic relaxation time. The other parameters are chosen as in Fig.~\ref{SD_BCF_plot}. 
    }
    \label{pb_diff_methods}
\end{figure}
In the weak-coupling regime, one often employs the Redfield master equation as a perturbative approximation of the atom dynamics, retaining only the second-order term in the Dyson expansion of the generator. It has the same form as Eq.~\eqref{eq:exact_me_GA}, but with the Lamb shift and decay rate given by their perturbative expressions
\begin{eqnarray}
   h(t) &=&   \mathrm{Im}\left[\Gamma_-(t)\right], \quad
    \gamma_-(t) = \mathrm{Re}\left[\Gamma_-(t)\right], \nonumber \\
   && \text{with} ~\Gamma_-(t) = \int_0^t ds\, C(s) e^{ i\omega_0 s};
\end{eqnarray}
see App.~\ref{app:TLS_GA_redfield_ME_zero_temp} for details. 
The Redfield master equation is more general than the Lindblad master equation, as it can account for bath memory effects that the Lindblad formalism inherently neglects. It has been shown to provide the most accurate description of the two–spin-boson model and the heat transport between two thermal baths mediated by two coupled oscillators in the weak-coupling regime~\cite{Hartmann2020,Scali2021}. 
However, as we demonstrate below, the Redfield approach may still fail even at weak coupling strengths, when the propagation delay $\tau$ becomes comparable to or exceeds the atomic relaxation time. 
The latter is defined as the inverse spontaneous emission rate in the Markovian limit at zero temperature, $\gamma_-^{-1}=[2\gamma(\omega_0)\cos^{2}{(\omega_0\tau/2)}]^{-1}$.

Figure~\ref{pb_diff_methods} shows the time evolution of the excitation probability of a two-level giant atom, $P_{ee}=\la e|\rho_S|e\ra$, at zero temperature and for three chosen time delay parameters $\tau$. The atom is initially prepared in the excited state. We compare the exact solution under the RWA (blue dotted), the Redfield master equation (black solid), and the HEOM simulation (red dashed).
Concerning the QuTiP HEOM implementation, the real and imaginary parts of the BCF at zero temperature are expanded into exponential terms using the ESPRIT method, as illustrated in Fig.~\ref{SD_BCF_plot}(b).
To ensure convergence of the HEOM simulation, we increased the number of ESPRIT fit terms from $(N_{\rm R},N_{\rm I}) = (6,7)$ in (a) to $(375,378)$ in (b) and $(743,750)$ in (c), while keeping the simulation depth fixed at $N_c=2$. 

In the zero-delay limit $\tau \rightarrow 0$, corresponding to the Markovian scenario shown in Fig.~\ref{pb_diff_methods}(a), the excitation probability exhibits expected exponential decay, with all three methods showing excellent agreement.
In contrast, the non-Markovian regime, in which the delay time (b) is comparable to or (c) exceeds the atomic relaxation time, leads to clear deviations from exponential decay. The Redfield master equation is then no longer accurate, and in particular, it fails to capture any excitation revivals in case (c), 
underscoring its limitations in modeling long-time memory effects. The numerically exact simulations on the other hand agree well with the RWA single-photon analytical results. HEOM is therefore well-suited for extensions to finite temperatures and stronger couplings, where an analytical solution based on the single-phonon assumption is no longer available. 

\section{Multi-phonon regime}\label{sec:multiphonon_regime}
We now consider the giant atom in a regime in which multi-phonon processes become significant, either due to the presence of thermal excitations of the SAW field or due to the ability of atomic transitions to excite multiple phonons. Such conditions could be achieved experimentally at higher temperatures and interaction strengths \cite{Anton2021,Chu2017}.
For GHz-frequency qubits typically operated in dilution refrigerators at temperatures of 10-100 mK, the ratio between the qubit transition energy and the thermal energy can be tuned from the deep quantum regime to a regime in which thermal excitations become appreciable. In addition, strong qubit–phonon coupling can be engineered by increasing the number of transducer finger pairs or by employing strongly piezoelectric substrates \cite{Chu2017}. 

We begin by simulating the effect of a finite temperature on the giant atom dynamics at weak coupling using HEOM and examine how thermal fluctuations influence the population of ground and excited state and the coherence between them. The temperature enters via the Ohmic BCF \eqref{eq:bcf_GA_OhmicSD}, which we fit with an adequate number of exponential functions, as we did at zero temperature.

\begin{figure}[t!]
    \centering
        \includegraphics[width=1\linewidth]{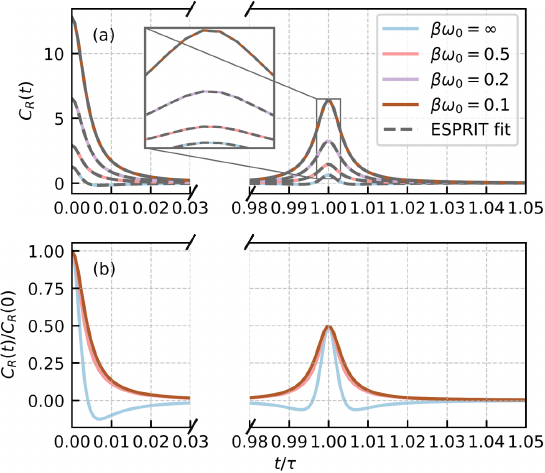}
    \caption{(a) Real part of the thermal BCF for the SAW field as a function of time at various temperatures, along with their corresponding ESPRIT fits (dashed). The inset provides a close-up of the second peaks and their fits.} (b) Rescaled thermal BCFs. All other parameters are the same as in Fig.~\ref{SD_BCF_plot}. 
    \label{fig:thermal_BCF_plot}
\end{figure}

The thermal behaviour and the quality of the fit are illustrated in Fig.~\ref{fig:thermal_BCF_plot}(a), which depicts the real part of the BCF $C_\mathrm{R}(t)$ as a function of time for various inverse temperatures $\beta \omega_0$, including the zero-temperature case $\beta \to \infty$.
In contrast, the imaginary part of the correlation function is temperature-independent and coincides with its zero-temperature form.
First, we notice that the overall amplitude of the BCF grows with temperature, scaling roughly in proportion to $1/\beta\omega_0$ when $\beta\omega_0 \ll 1$. 
To highlight the temperature dependence of the $C_R(t)$ shape, Fig.~\ref{fig:thermal_BCF_plot}(b) presents the normalized curves $C_R(t)/C_R(0)$. 
The shape of the BCF also changes: the peaks around $t=0$ and $\tau$ broaden with growing temperature and, unlike at zero temperature [see Fig.~\ref{SD_BCF_plot}(b)], the peak tails are no longer negative for $\beta\omega_0 < 2$. The peak ratio stays at $C_\mathrm{R}(0)/C_\mathrm{R} (\tau) \approx 2$ as long as the memory time is large, $\omega_0 \tau \gg 1$. 
The dashed lines in Fig.~\ref{fig:thermal_BCF_plot}(a) show the ESPRIT fits, which replicate $C_{\mathrm{R}}(t)$ accurately across all temperatures. As the structure of the BCF simplifies with growing temperature, the number of exponential fit terms can be reduced. The four plotted fits were obtained with $(N_{\rm R},N_{\rm I}) = (743,750)$, $(645,750)$, $(573,750)$, and $(545,750)$ in order of decreasing $\beta$. 

\begin{figure}[t!]
    \centering
    \includegraphics[width=\linewidth]{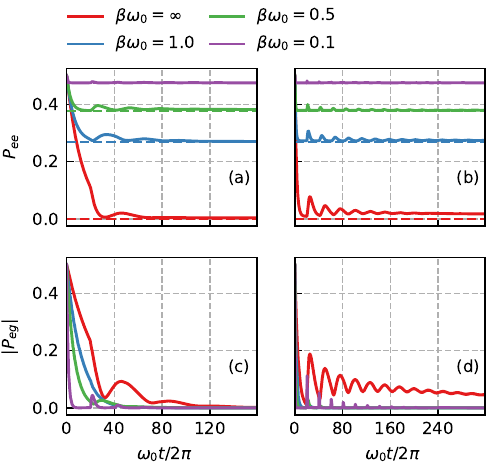}
    \caption{Time evolution of the excited-state population $P_{ee}$ [panels (a), (b)] and coherence $|P_{eg}|$ [panels (c), (d)] of a two-level giant atom at different temperatures with a hierarchy depth $N_c =2$. 
    Results are shown for weak coupling ($\eta = 1\times10^{-2}$, top 2) and stronger coupling ($\eta = 5\times 10^{-2}$, bottom 2). 
    All other parameters are the same as in Fig.~\ref{SD_BCF_plot}. The dashed lines indicate the corresponding Gibbs-state values.}
    \label{fig:pb_coh_diff_eta}
\end{figure}

With the BCF fits at hand, we can simulate the evolution of the two-level giant atom using HEOM  with a depth of $N_c=2$. Fig.~\ref{fig:pb_coh_diff_eta}(a) and (c) show the time evolution of the excited state population, $P_{ee}(t) = \la e|\rho_S (t)|e\ra$, and the coherence, $P_{eg}(t) = \la e|\rho_S(t)|g\ra$, respectively, for a fairly weak coupling strength $\eta = 10^{-2}$. The atom is initialized in the superposition state $|+\ra = (|g\ra + |e\ra)/\sqrt{2}$. In (a), the dashed lines mark the Gibbs steady-state population for each temperature, which is reached at long times. In (c), the coherence decays more rapidly with growing temperature. Nevertheless, revivals in both population and coherence persist, albeit with reduced amplitude and faster damping. This resilience of the memory effect against thermal noise constitutes a key feature of giant atom systems, which lends itself to benchmarking numerical or approximate techniques for non-Markovian open systems.

We now turn to the role of stronger coupling. The atom-bath coupling strength $\eta$ enters the BCF \eqref{eq:bcf_GA_OhmicSD} merely as a prefactor. Greater $\eta$-values thus result in a higher number of more pronounced revivals in both excitation and coherence, but with a faster decay of each such revival. This is illustrated in Fig.~\ref{fig:pb_coh_diff_eta}(b) and (d), which display the evolution of both quantities for a five times stronger coupling than in (a) and (c). 
Overall, these results indicate that stronger coupling enhances non-Markovian memory effects by reinforcing the time-delayed feedback between the system and its environment.
Remarkably, at zero temperature, we not only observe the clearest non-Markovian features, but also the onset of systematic deviations from the equilibration to a diagonal Gibbs state with respect to the bare atom Hamiltonian. 
At finite temperatures, the steady-state populations still deviate slightly from the corresponding Gibbs values (dashed lines), while the coherences vanish in the parameter regimes considered here. Thus, although delayed feedback produces pronounced transient non-Markovian revivals, it does not by itself generate steady coherence in the present two-level giant-atom setup. Steady-state coherence may nevertheless be engineered in suitably designed environment, for example through tailored system-bath coupling structures \cite{Guarnieri2018, Purkayastha2020}.
The observed population deviations are consistent with the expectation that equilibrium properties are modified by system–bath correlations and can be described by the Hamiltonian of mean force \cite{Hanggi2009,Rivas2020}. For our moderate coupling strengths, we could confirm this behavior by  computing the mean-force Hamiltonian using a perturbative expansion; see App.~\ref{app:cal_mean_force_Hamiltonian} for details.
Table~\ref{tab:thermal_population} compares the excited-state populations of the giant atom predicted by the second-order mean-force Hamiltonian ($P_{ee}^*$) with those obtained from HEOM ($P_{ee}$) in the stationary regime, as well as with the Gibbs-state population of the bare system Hamiltonian ($P_{ee}^{th}$) at various temperatures. At finite temperatures, the second-order mean-force Hamiltonian yields excited-state populations that almost exactly coincide with the HEOM results, whereas  the corresponding Gibbs-state values exhibit a small deviation.
In the zero-temperature limit, the second-order mean-force Hamiltonian no longer provides quantitatively reliable predictions.

\begin{figure}[t!]
    \centering
    \includegraphics[width=\linewidth]{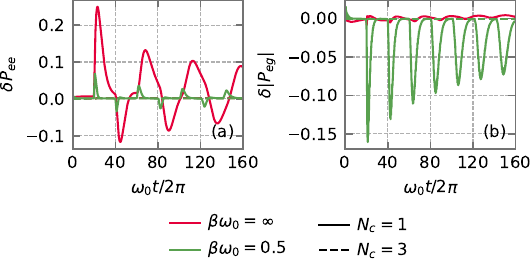}
    \caption{HEOM convergence with hierarchy depth for $\eta$=0.05. The differences of (a) $P_{ee}$ and (b) $|P_{eg}|$ between our used depth $N_c =2$ and the depth values $N_c = 1$ (solid) and 3 (dashed) are plotted as functions of time for $\beta \omega_0=\infty$ and $0.5$. The dashed curves do not noticeably deviate from zero. All other parameters are the same as in Fig.~\ref{fig:pb_coh_diff_eta}.}
    \label{fig:pb_coh_diff_depth}
\end{figure}

To ensure that the observed features are not artifacts of the HEOM truncation at $N_c=2$, we performed an explicit convergence analysis with respect to the hierarchy depth. This test was carried out for the stronger-coupling case $\eta=0.05$, where the influence of higher-tier auxiliary density operators is expected to be more relevant due to stronger system--environment correlations. The comparison shown in Fig.~\ref{fig:pb_coh_diff_depth} was performed at both zero temperature, $\beta\omega_0=\infty$, and high temperature, $\beta\omega_0=0.5$. In both cases, the results obtained with $N_c=2$ and $N_c=3$ are practically indistinguishable, whereas the shallow hierarchy with $N_c=1$ leads to significant deviations and even unphysical behavior (i.e., $P_{ee}<0$). 
Regarding computational cost, we note that the number of ADOs used in the HEOM \eqref{eq:heom_eqs} grows combinatorially with both the hierarchy depth $N_c$ and the total number of exponential terms $K = N_R + N_I$ used in the decomposition of the BCF,
\begin{equation}
N_{\mathrm{ADO}} = \frac{(K + N_c)!}{K! \, N_c!}.
\end{equation}
This scaling leads to a rapid increase in both memory usage and computational time as $N_c$ increases. In the present system, the number of exponential terms $K$ is already moderately large due to the structured BCF and its ESPRIT-based decomposition.
Extending the hierarchy to, say, $N_c = 4$ would result in a substantial increase in the number of ADOs, making the simulations significantly more demanding in terms of both memory and runtime, even on high-performance computing platforms. Since we observed no appreciable changes from $N_c = 2$ to $N_c = 3$, we conclude that $N_c = 2$ is sufficient to ensure numerical convergence. However, we note that in regimes of stronger system-bath coupling, higher hierarchy depths may become necessary. In such cases, the computational cost could be mitigated by considering shorter time delays, which decreases the number of exponential terms required to represent the BCF and thus the total number of ADOs. 
Our choice of a relatively long time delay is deliberate, as it enables us to clearly observe non-Markovian memory effects even in the weak-to-moderate coupling regime, without the need to invoke ultrastrong coupling. And it is this regime that was accessed experimentally in Ref.~\cite{Andersson2019}.

\begin{table}[t]
    \centering
     \small
    \setlength{\tabcolsep}{6pt}
    \renewcommand{\arraystretch}{1.15}
    \begin{tabular}{|c|c|c| c| c|}
        \hline
         & $\beta \omega_0 = \infty$
        & $\beta \omega_0 = 1.0$
        & $\beta \omega_0 = 0.5$
         & $\beta \omega_0 = 0.1$ \\ \hline
         $P_{ee}$ & 0.009 & 0.272 & 0.378 & 0.475 \\ \hline
         $P_{ee}^*$ & 0 & 0.271 & 0.378 & 0.475 \\ \hline
        $P_{ee}^{(th)}$ & 0 & 0.269 & 0.377 & 0.475 \\ \hline
    \end{tabular}
    \caption{Comparison of the stationary excited-state populations obtained from HEOM ($P_{ee}$), the Hamiltonian of mean force ($P_{ee}^*$), and the bare system Hamiltonian  ($P_{ee}^{(th)}$) at several temperatures. Please refer to Fig.~\ref{fig:pb_coh_diff_eta}(b) for the parameter settings.}
    \label{tab:thermal_population}
\end{table}

\section{Conclusion}\label{sec:conclusion}
We have developed a numerically exact framework for giant-atom dynamics by combining HEOM with an optimized ESPRIT-based representation of structured bath correlations. This enables access to regimes with long memory, finite temperature, and strong coupling. We demonstrate the breakdown of perturbative non-Markovian approaches in the presence of delay-induced memory, and show that non-Markovian effects remain robust against thermal fluctuations and can even be enhanced at stronger coupling.

Our results establish a general and efficient framework for studying non-Markovian dynamics in structured environments, with direct relevance for quantum technologies based on giant atoms. More broadly, this approach provides a route to quantitatively analyze memory effects and engineer environments in emerging quantum acoustic and waveguide QED platforms \cite{Andersson2019,Grimsmo2015,Pichler2016}.
Future extensions include the study of multi–contact-point giant atoms, where the system–bath coupling acquires multiple phase factors associated with different coupling points, leading to increasingly structured bath correlation functions and richer feedback mechanisms. Such scenarios can be naturally incorporated within the present framework, as they primarily modify the structure of the BCF without altering the underlying HEOM formalism.
Additional directions include collective emission in multi-atom networks, interactions with more complex structured environments, and driven nonequilibrium dynamics \cite{Li2024,Ritter2025}. It would also be interesting to investigate how memory effects in giant-atom systems persist at finite temperature, building on recent theoretical frameworks \cite{Backer2024,Yu2025}. These directions may further advance the development of controllable non-Markovian platforms for quantum information processing and quantum thermodynamics.

\acknowledgments
M.Y.\ would like to thank Valentin Link and
Hai-Chau Nguyen for helpful discussions. The University of Siegen is kindly acknowledged for enabling our computation through the OMNI cluster.

\onecolumngrid
\appendix

\section{Exact master equation in RWA at zero temperature}
\label{app:TLS_GA_exa_ME_zero_temp}
Here we derive the master equation \eqref{eq:exact_me_GA} based on the exact solution for the evolution of the giant atom and the SAW field under the RWA at zero temperature. 
The RWA allows us to drop the counter-rotating interaction terms in the atom-bath Hamiltonian \eqref{total_hamiltonian}, so that in the interaction picture with respect to $\oH_0 =  \hbar \omega_0 |e\rangle \langle e| + \hbar \sum_{j=R,L} \int_0^{\infty} d\omega \omega \hat{a}_{j\omega}^{\dagger} \hat{a}_{j\omega}$, the interaction Hamiltonian reduces to 
\begin{eqnarray} \label{Hamiltonian_RWA_INT}
   \oH_{I}(t) = \hbar \sum_{j=R,L}\int_0^{\infty} d\omega \sqrt{\frac{\gamma(\omega)}{\pi}}\cos \left(\frac{\omega \tau}{2} \right) \left[ \oa_{j\omega}^{\dagger} e^{-i\left(\omega_0 -\omega\right) t} \osigma_- + \oa_{j\omega} e^{i\left(\omega_0 - \omega\right) t} \osigma_+ \right] . \label{int_hamiltonian_two_level}
\end{eqnarray}
Given that the phonon field is initially in the vacuum state and the total excitation number of atom and field is conserved, there can be at most a single phonon excitation present at any time $t$. A global pure state can thus be described by the ansatz
\begin{eqnarray}
|\Psi(t) \rangle = \sum_{j=R,L}\int d\omega \, \alpha_{j\omega}(t) a_{j\omega}^{\dagger} |g,0\rangle + e(t) |e,0\rangle + g(0) |g, 0\rangle ,
\end{eqnarray}
with the initial condition $\alpha_{\omega} (0) = 0$. The amplitude of the $|g,0\rangle$ component remains constant in time, as $\oH_{I}(t) |g, 0\rangle = 0$.
Using the Schrödinger equation $i\hbar \partial_t |\Psi(t) \rangle = \oH_{I}(t) |\Psi(t) \rangle$, we obtain the coupled equations of motion for the probability amplitudes
\begin{eqnarray}
    \dot{e}(t) &=& -i \sum_{j=1,2} \int_0^{\infty} d\omega \sqrt{\frac{\gamma(\omega)}{\pi}} \cos{\left(\frac{\omega \tau}{2}\right)} \alpha_{j\omega}(t) e^{i\left(\omega_0 -\omega\right)t} \label{EOM_sys}, \\
    \dot{\alpha}_{j\omega}(t) &=&  -i \sqrt{\frac{ \gamma(\omega)}{\pi}} \cos{\left(\frac{\omega \tau}{2}\right)} e(t) e^{-i\left(\omega_0 - \omega \right)t}. \label{EOM_env}
\end{eqnarray}
By formally solving $\alpha_{j\omega}(t)$ and substituting the result into Eq.~\eqref{EOM_sys}, we obtain an integro-differential equation for the amplitude $e(t)$. Writing $e(t) = G(t) e(0)$ in terms of a propagator function $G(t)$, the equation can be stated as
\begin{eqnarray}\label{G_diff_funct_app}
    \dot{G}(t) &=& -\int_0^t ds\, C(s)G(t-s)e^{i\omega_0 s},
\end{eqnarray}
with initial condition $G(0)=1$. Here, $C(s)$ is the BCF of the phonon field at zero temperature, 
\begin{equation} \label{bcf_zero_temp_GA}
    C(s) = \frac{1}{\pi} \int_0^{\infty} d\omega\, 2\gamma(\omega) \cos^2{\left(\frac{\omega \tau}{2}\right)} e^{-i\omega s}.
\end{equation}
The solution of $G(t)$ can be found through a Laplace transform or a Fourier transform with appropriate boundary conditions \cite{Guo2017}. 
From this, the reduced density matrix of the giant atom in the interaction picture can be obtained by tracing over the field, 
\begin{eqnarray}
    \tilde{\rho}_S(t) &=& \tr_f\{|\Psi(t)\rangle \langle \Psi(t)|\} = 
    \begin{bmatrix}\rho_{ee}(t) & \rho_{eg}(t)
    \\ \rho_{ge}(t) & \rho_{gg}(t) \end{bmatrix} = \begin{bmatrix}|G(t)|^2\rho_{ee}(0) & G(t)\rho_{eg}(0)
    \\ G^*(t)\rho_{eg}^*(0) & 1-|G(t)|^2\rho_{ee}(0) \end{bmatrix}. \label{dyn_map}
\end{eqnarray}
This describes how the populations and the coherence of the giant atom evolve in time, governed by the function $G(t)$.
It is important to note that, although these equations have been derived assuming a pure initial product state, they are also valid for any mixed initial state of the giant atom. This generality arises from the fact that a mixed state can be represented as a convex linear combination of pure states, and the function $G(t)$ is independent of the initial condition.

To arrive at a generator of the exact time evolution, we can simply take the time derivative of \eqref{dyn_map} and express it in terms of a linear superoperator, $d \tilde{\rho}_S(t)/dt = \mathcal{K}(t) \tilde{\rho}_S(t)$. After a few steps of algebra, we find the closed form,
\begin{eqnarray}
  \mathcal{K}(t) \tilde{\rho}_S(t) 
    &=& -i h(t) \left[\osigma_+ \osigma_-, \tilde{\rho}_S(t)\right] + \gamma_{-}(t)\mathcal{D}\left[\osigma_-\right]\tilde{\rho}_S(t).
    \label{ex_NM_qme}
\end{eqnarray}
Here, $h(t) = -\mathrm{Im} [\dot{G}(t)/G(t)]$ is the Lamb shift of the system energy and $\mathcal{D}[\osigma_-]\tilde{\rho}_S(t) = 2 \osigma_- \tilde{\rho}_S(t)\osigma_+ - \{\osigma_+ \osigma_-,\tilde{\rho}_S(t)\}$ is the dissipator describing spontaneous emission with a time-dependent decay rate $\gamma_-(t) = -\mathrm{Re}[\dot{G}(t)/G(t)]$. Transforming back to the Schrödinger picture, we obtain the master equation \eqref{eq:exact_me_GA} in the main text.

\section{Redfield master equation in RWA at zero temperature}
\label{app:TLS_GA_redfield_ME_zero_temp}
Here we derive the explicit form of the Redfield master equation for the giant atom under RWA at zero temperature, which is compared to the exact solution in the main text. The Redfield master equation is a time-local differential equation for the reduced state of an open quantum system that describes its evolution to lowest perturbative order in the weak coupling to a large reservoir \cite{Breuer2002}. In the interaction picture, it is given by
\begin{equation}
    \frac{d}{dt} \tilde{\rho}_S(t) =- \frac{1}{\hbar^2}\int_{0}^t ds\, \tr_B \left\{ \big[  \oH_I(t), \big[  \oH_I(s),  \tilde{\rho}_S(t) \otimes \rho_B(0)\big]\big] \right\}. \label{redfield_ME_gene}
\end{equation}
Inserting the interaction-picture Hamiltonian for our case, $\oH_I (t) = \hbar \tilde{\sigma}_x (t) \tilde{B}(t)$ with $\osigma_x(t) = \osigma_- e^{-i\omega_0 t} + h.c.$ and $\oB(t)$ given in \eqref{eq:B}, the master equation can be expressed as  
\begin{equation}
     \frac{d}{dt}  \tilde{\rho}_S(t) = - \int_{0}^t dt'\left\{ C(t-t') \big[\osigma_x(t), \osigma_x(t') \tilde{\rho}_S(t)\big] + \mathrm{h.c.}\right\}. \label{redfield_me_bcf}
\end{equation}
Here, the properties of the phonon bath are encoded in the BCF $C(t)$. Assuming a thermal initial state $\rho_B(0)$ at inverse temperature $\beta = \hbar/k_{\rm B}T$, the BCF reads as
\begin{eqnarray} \label{bcf_GA_finite_temp}
    C(s) &=& \langle \tilde{B}(t)  \tilde{B}(t+s)\rangle = \sum_{j=R,L} \int_0^{\infty} d\omega \frac{\gamma(\omega)}{\pi} \cos^2{\left(\frac{\omega \tau}{2}\right)} \nonumber  \left[ \frac{2\cos{\omega s}}{e^{\beta \omega}-1} 
    +e^{-i\omega s}\right] = \frac{1}{\pi} \int_{-\infty}^{\infty} d\omega \, J(\beta, \omega) e^{-i\omega s}, 
\end{eqnarray}
introducing the effective bath spectral density
\begin{eqnarray}
    J(\beta, \omega) := \frac{2\gamma(\abs{\omega})}{\abs{1-e^{-\beta \omega}}} \cos^2{\left(\frac{\omega \tau}{2}\right)}.
\end{eqnarray}
Transforming back to the Schrödinger picture, the master equation becomes
\begin{eqnarray} \label{redfield_local_me_GA}
   \frac{d}{dt} \rho_S(t) = -i \omega_0 \big[\osigma_+ \osigma_-, \rho_{S}(t)\big] - \int_0^{t} ds \{ C(s) \big[ \osigma_x, \osigma_x(-s) \rho_{S}(t)\big] + \mathrm{h.c.}\},
\end{eqnarray}
where the $\osigma_x$ and $\osigma_{\pm}$ without time argument refer to the Schrödinger-picture operators. 
Finally, we can expand the remeining $\osigma_x(-s) = \osigma_-e^{i\omega_0 s} + \osigma_+e^{-i\omega_0 s}$, which leads to 
\begin{equation} \label{redfield_local_me_GA_sim}
   \frac{d}{dt} \rho_S(t)  = -i \omega_0[\osigma_+\osigma_-, \rho_S(t)] +  \big\{\Gamma_+(t)[\osigma_+\rho_S(t), \osigma_x]  + \mathrm{h.c.} \big\} + \big\{\Gamma_-(t)[\osigma_-\rho_S(t), \osigma_x] + \mathrm{h.c.} \big\},
\end{equation}
with the time-dependent coefficients $\Gamma_{\pm}(t) = \int_0^t ds C(s) e^{\mp i\omega_0 s}$.
Under the RWA, we drop all counter-rotating terms and keep only those operator products containing both $\osigma_+$ and $\osigma_-$, resulting in 
\begin{eqnarray}
   \frac{d}{dt} \rho_S(t) 
    \approx  -i\left[\omega_0 \osigma_+\osigma_- + \mathrm{Im}\left[\Gamma_+(t)\right]\osigma_-\osigma_+ + \mathrm{Im}\left[\Gamma_-(t)\right]\osigma_+ \osigma_-, \rho_S(t)\right] + \mathrm{Re}\left[\Gamma_+(t)\right]\mathcal{D}[\osigma_+]\rho_S(t) + \mathrm{Re}\left[\Gamma_-(t)\right]\mathcal{D}[\osigma_-]\rho_S(t). \nonumber \\
\end{eqnarray}
In this form, it is evident that the real and imaginary parts of $\Gamma_\pm(t)$ determine the dissipative and coherent contributions, respectively.
At zero temperature, the coefficient $\Gamma_+(t)$ becomes strongly suppressed compared to $\Gamma_-(t)$ and can thus be neglected. We are left with
\begin{eqnarray} \label{GA_RF_me}
     \frac{d}{dt} \rho_S(t)=  -i\left(\omega_0 + \mathrm{Im}\left[\Gamma_-(t)\right]\right)[\osigma_+\osigma_-, \rho_S(t)] + \mathrm{Re}\left[\Gamma_-(t)\right]\mathcal{D}[\hat{\sigma}_-]\rho_S(t),
\end{eqnarray}
which we use for comparison with our exact results in Sec.~\ref{sec:benchmark} of the main text.

\section{Dependence on  the exponential cutoff frequency}
\label{app:dependence_cutoff_freqs.}
Varying the cutoff frequency $\omega_c$ modifies not only the memory structure of the BCF, but also the effective system--bath coupling strength. To isolate the genuine cutoff-induced memory effects on the system dynamics from the trivial increase of the overall bath strength with $\omega_c$, we renormalize the coupling parameter in the Ohmic spectral density at zero temperature according to
\begin{equation}
\eta \rightarrow \frac{\eta(1+\omega_c^2\tau^2)}{(2+\omega_c^2\tau^2)\omega_c},
\end{equation}
such that the effective reorganization energy
$\int_0^{\infty} d\omega J_r(\omega)/\omega$
remains constant for different cutoff frequencies \cite{Cheng2009, Ritschel2011}. Here, the reorganization energy characterizes the overall effective system-bath coupling strength experienced by the giant atom. Consequently, both the spectral density and the corresponding BCF are rescaled by the same renormalization factor $
(1+\omega_c^2\tau^2)/{(2+\omega_c^2\tau^2)\omega_c}$.
Although this renormalization is derived at zero temperature, the same scaling is also adopted at finite temperature, since thermal effects modify the occupation factor in the effective bath spectral density or BCF but do not alter the underlying coupling strength.
 
\begin{figure}[t!]
    \centering
        \includegraphics[width=1\linewidth]{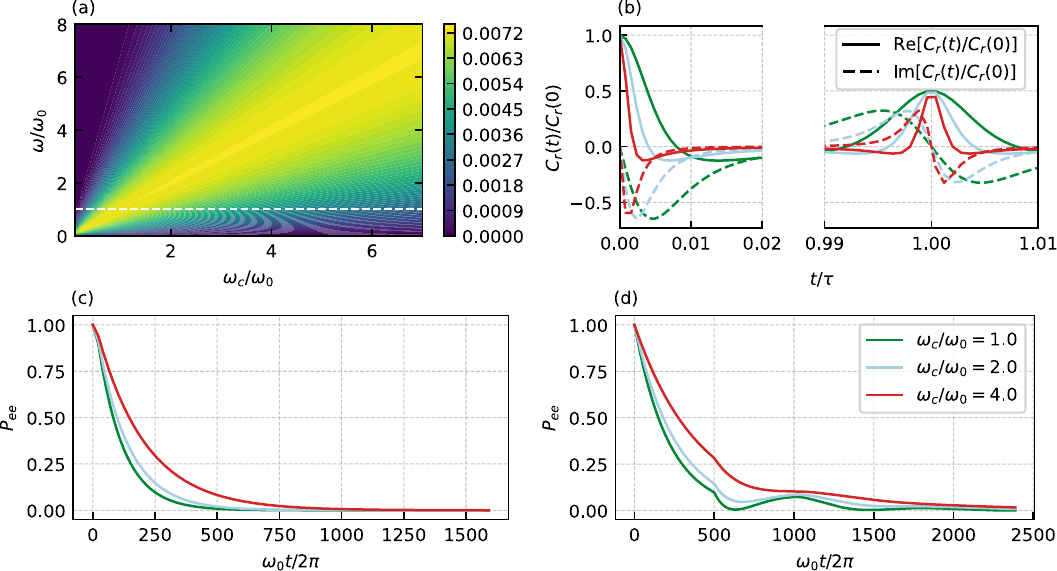}
    \caption{(a)Renormalized bath spectral density $J_r(\omega, \omega_c)$ at zero temperature. (b) Rescaled BCFs for different cutoff frequencies. The corresponding excited-state probability for (a) delay time $\omega_0\tau/2\pi = 20$ and (b) $\omega_0\tau/2\pi = 500$. All other parameters are the same as in Fig.~\ref{SD_BCF_plot}}. 
    \label{fig:cutoff_eff_plot}
\end{figure}

As shown in Fig. \ref{fig:cutoff_eff_plot}(a), varying $\omega_c$ reshapes the spectral distribution of the bath. When the cutoff frequency becomes comparable to the system resonance frequency, the spectral weight around the resonance region is enhanced, leading to stronger damping in the system dynamics.
Fig. \ref{fig:cutoff_eff_plot}(b) presents the corresponding normalized BCF $C_r(t)/C_r(0)$, which isolates the temporal memory structure of the bath independently of the overall correlation amplitude. Smaller cutoff frequencies lead to more slowly decaying bath correlations, corresponding to longer bath memory times and thus stronger non-Markovian effects, whereas larger cutoff frequencies produce more rapidly decaying correlations and shorter memory times.
The corresponding excited-state population dynamics shown in Fig. \ref{fig:cutoff_eff_plot}(c)-(d) for different delay time demonstrate the combined influence of the spectral redistribution and the bath memory effects induced by the cutoff frequency. The corresponding excited-state population dynamics shown in Fig.~\ref{fig:cutoff_eff_plot}(c)-(d) demonstrate the combined influence of spectral redistribution and bath memory effects induced by the cutoff frequency. We consider two different delay times. In the first case, the delay time is much shorter than the corresponding renormalized atomic relaxation timescale, such that delayed feedback returns only after the excitation has already substantially decayed, resulting in the absence of pronounced non-Markovian revivals. In the second case, the delay time becomes comparable to the effective atomic relaxation timescale, allowing the delayed feedback to interfere significantly with the system dynamics and thereby generating clear revival structures. Since the effective relaxation timescale itself depends on the cutoff frequency through the renormalized spectral distribution, both the strength and timing of the revivals vary with $\omega_c$. Note that the bath memory time is determined by the decay of the bath correlation function, whereas the population relaxation rate is controlled by the spectral weight near the system resonance, i.e., $J_r(\omega_0,\omega_c)$. Therefore, a smaller cutoff frequency may lead to both longer memory times and faster population decay if it enhances the effective spectral density around the resonance frequency.

\section{Calculation of mean-force Hamiltonian} \label{app:cal_mean_force_Hamiltonian}
Here we present a perturbative method for evaluating the Hamiltonian of mean force. From this, the corresponding Gibbs state can be constructed and compared with the stationary state obtained numerically via HEOM in Sec.~\ref{sec:multiphonon_regime}. The Hamiltonian of mean force is defined as \cite{Hanggi2009,Rivas2020}
\begin{equation}
    \oH_S^{*}  = -\frac{1}{\tilde{\beta}}\log{\bigg[\frac{\mathrm{tr}_B\{e^{-\tilde{\beta} \oH}\}}{\mathcal{Z_B}}}\bigg],
\end{equation}
where $\mathcal{Z}_B$ is partition function of the bath and  here $\tilde{\beta} = \beta/\hbar = 1/k_B T$. To proceed, we apply the imaginary-time Dyson identity \cite{Lobejko2024}
\begin{eqnarray}
    e^{-\tilde{\beta} \oH} = e^{-\tilde{\beta} \left(\oH_S + \oH_B \right)} \mathcal{T} e^{-\int_0^{\tilde{\beta}} d\tilde{t}\oH_I(\tilde{t})},
\end{eqnarray}
with the shorthand notation $\oH_I(\tilde{t}) = e^{\tilde{t}(\oH_S+ \oH_B)} \oH_I e^{\tilde{t}(\oH_S+ \oH_B)}$ and $\mathcal{T}$ is the time-ordering operator. Notice that the imaginary time variable $\tilde{t}$ is now given in units of inverse energy.
Using the identity, the Hamiltonian of mean force can be written in the Dyson-expanded form
\begin{eqnarray}
    \oH_S^{*}  &=& -\frac{1}{\tilde{\beta}} \log \Big[e^{-\tilde{\beta} H_s} \la \mathcal{T} e^{-\int_0^{\tilde{\beta}} d\tilde{t}'\oH_I(\tilde{t}')}\ra_B \Big]  = \oH_S -\frac{1}{\tilde{\beta}} \log \la \mathcal{T} e^{-\int_0^{\tilde{\beta}} d\tilde{t}'\oH_I(\tilde{t}')}\ra_B,
\end{eqnarray}
where $\langle\,\cdot\,\rangle_B$ denotes the thermal expectation value over the bath and 
\begin{eqnarray}
     \la \mathcal{T} e^{-\int_0^{\tilde{\beta}} d\tilde{t}'\oH_I(\tilde{t}')}\ra_B = 1 - \int_0^{\tilde{\beta}} d\tilde{t} \la \oH_I(\tilde{t}) \ra_B + \int_0^{\tilde{\beta}} d\tilde{t}_1 \int_0^{\tilde{t}_1} d\tilde{t}_2 \la \oH_I(\tilde{t}_1) \oH_I(\tilde{t}_2)\ra_B + \ldots
\end{eqnarray}
The first-order contribution $\langle \oH_I(\tilde{t})\rangle_B$ typically vanishes, and does so in our case. For moderate coupling strengths, a second-order truncation provides an accurate approximation. We will show that this approximation is sufficient for the coupling regime considered in Fig.~\ref{fig:pb_coh_diff_methods}(b). 
Consider the interaction between the giant atom and SAWs in Eq.~\eqref{total_hamiltonian}, the second-order term is given by
\begin{eqnarray}
 \int_0^{\tilde{\beta}} d\tilde{t}_1 \int_0^{\tilde{t}_1} d\tilde{t}_2 \la \oH_I(\tilde{t}_1) \oH_I(\tilde{t}_2)\ra_B &=&I_1(\tilde{\beta}) \osigma_+\osigma_- + I_2(\tilde{\beta}), \quad \text{with} \nonumber \\
  I_1(\tilde{\beta}) &=&  \hbar^2\int_0^{\tilde{\beta}} d\tilde{t}_1 \int_0^{\tilde{t}_1} d\tilde{s} \, \tilde{C}(\tilde{s}) \sinh{\hbar\omega_0 \tilde{s}} ,  \nonumber \\
  I_2(\tilde{\beta}) &=& \hbar^2 \int_0^{\tilde{\beta}} d\tilde{t}_1 \int_0^{\tilde{t}_1} d\tilde{s} \, \tilde{C}(\tilde{s})\big[\cosh{\hbar\omega_0(2\tilde{t}_1-\tilde{s})} + e^{-\hbar \omega_0 \tilde{s}} \big].
\end{eqnarray}
Here, $\tilde{C}(s)$ is the BCF in the imaginary-time interaction picture, which for our Ohmic setting is given by
\begin{eqnarray}
   \tilde{C}(\tilde{s}) &=& \frac{\eta}{2\pi \hbar^2 \tilde{\beta}^2} \Biggl\{2\Psi\bigg[1,\frac{1+ \tilde{s} \hbar \omega_c}{\tilde{\beta} \hbar \omega_c}\bigg] + 2\Psi \bigg[1,1 - \frac{\tilde{s}}{\tilde{\beta}} + \frac{1}{\tilde{\beta} \hbar \omega_c}\bigg]  
    + \Psi \bigg[1,\frac{1+ \tilde{s} \hbar \omega_c -i \tau \omega_c}{\tilde{\beta} \hbar \omega_c}\bigg] + \Psi \bigg[1,\frac{1 + \tilde{s} \hbar \omega_c +i\tau\omega_c}{\tilde{\beta} \hbar \omega_c}\bigg] \nonumber \\
    &+&\Psi \bigg[1,\frac{1+\left(\tilde{\beta} - \tilde{s} \right) \hbar \omega_c  - i\tau \omega_c}{\tilde{\beta} \hbar \omega_c}\bigg] + \Psi \bigg[1,\frac{1+\left(\tilde{\beta} - \tilde{s} \right) \hbar \omega_c  + i\tau \omega_c}{\tilde{\beta} \hbar \omega_c}\bigg].
\end{eqnarray}
We evaluate $I_1(\tilde{\beta})$ and $I_1(\tilde{\beta})$ numerically. Thus, the mean-force Hamiltonian to second order is given by
\begin{eqnarray}
    \oH_S^* &\approx& \oH_S + \frac{\delta \omega(\tilde{\beta})}{2} \osigma_z + c(\tilde{\beta}), \text{with} \nonumber \\
    \delta \omega(\tilde{\beta}) &=& -\frac{\log{[1+I_1(\tilde{\beta}) + I_2(\tilde{\beta})]} - \log{[1 + I_2(\tilde{\beta})]}}{\tilde{\beta}}, \nonumber \\
    c(\tilde{\beta}) &=&  -\frac{\log{[1+I_1(\tilde{\beta}) + I_2(\tilde{\beta})]} + \log{[1 + I_2(\tilde{\beta})]}}{2\tilde{\beta}}.
\end{eqnarray}
Note that the offset term $c(\tilde{\beta})$ does not affect the thermodynamical properties and therefore can be neglected.
The corresponding Gibbs state is $\rho_S^*=e^{-\tilde{\beta} \oH_S^*}/\mathcal{Z}_S^*$ with $\mathcal{Z}_S= \mathrm{tr}_S(e^{-\tilde{\beta}\oH_S^*})$. The excited-state probability is then obtained as
\begin{equation}
    P_{ee}^* = \la e|\rho_S^*|e\ra \approx \left(1+e^{\beta \omega_0 - \log{[1+I_1(\tilde{\beta}) + I_2(\tilde{\beta})]} + \log{[1+I_2(\tilde{\beta})]} } \right)^{-1}.
\end{equation}


\begin{thebibliography}{75}%
\makeatletter
\providecommand \@ifxundefined [1]{%
 \@ifx{#1\undefined}
}%
\providecommand \@ifnum [1]{%
 \ifnum #1\expandafter \@firstoftwo
 \else \expandafter \@secondoftwo
 \fi
}%
\providecommand \@ifx [1]{%
 \ifx #1\expandafter \@firstoftwo
 \else \expandafter \@secondoftwo
 \fi
}%
\providecommand \natexlab [1]{#1}%
\providecommand \enquote  [1]{``#1''}%
\providecommand \bibnamefont  [1]{#1}%
\providecommand \bibfnamefont [1]{#1}%
\providecommand \citenamefont [1]{#1}%
\providecommand \href@noop [0]{\@secondoftwo}%
\providecommand \href [0]{\begingroup \@sanitize@url \@href}%
\providecommand \@href[1]{\@@startlink{#1}\@@href}%
\providecommand \@@href[1]{\endgroup#1\@@endlink}%
\providecommand \@sanitize@url [0]{\catcode `\\12\catcode `\$12\catcode
  `\&12\catcode `\#12\catcode `\^12\catcode `\_12\catcode `\%12\relax}%
\providecommand \@@startlink[1]{}%
\providecommand \@@endlink[0]{}%
\providecommand \url  [0]{\begingroup\@sanitize@url \@url }%
\providecommand \@url [1]{\endgroup\@href {#1}{\urlprefix }}%
\providecommand \urlprefix  [0]{URL }%
\providecommand \Eprint [0]{\href }%
\providecommand \doibase [0]{https://doi.org/}%
\providecommand \selectlanguage [0]{\@gobble}%
\providecommand \bibinfo  [0]{\@secondoftwo}%
\providecommand \bibfield  [0]{\@secondoftwo}%
\providecommand \translation [1]{[#1]}%
\providecommand \BibitemOpen [0]{}%
\providecommand \bibitemStop [0]{}%
\providecommand \bibitemNoStop [0]{.\EOS\space}%
\providecommand \EOS [0]{\spacefactor3000\relax}%
\providecommand \BibitemShut  [1]{\csname bibitem#1\endcsname}%
\let\auto@bib@innerbib\@empty
\bibitem [{\citenamefont {Du}\ \emph {et~al.}(2022)\citenamefont {Du},
  \citenamefont {Zhang}, \citenamefont {Wu}, \citenamefont {Kockum},\ and\
  \citenamefont {Li}}]{Du2022}%
  \BibitemOpen
  \bibfield  {author} {\bibinfo {author} {\bibfnamefont {L.}~\bibnamefont
  {Du}}, \bibinfo {author} {\bibfnamefont {Y.}~\bibnamefont {Zhang}}, \bibinfo
  {author} {\bibfnamefont {J.-H.}\ \bibnamefont {Wu}}, \bibinfo {author}
  {\bibfnamefont {A.~F.}\ \bibnamefont {Kockum}},\ and\ \bibinfo {author}
  {\bibfnamefont {Y.}~\bibnamefont {Li}},\ }\bibfield  {title} {\bibinfo
  {title} {Giant atoms in a synthetic frequency dimension},\ }\href
  {https://doi.org/10.1103/PhysRevLett.128.223602} {\bibfield  {journal}
  {\bibinfo  {journal} {Phys. Rev. Lett.}\ }\textbf {\bibinfo {volume} {128}},\
  \bibinfo {pages} {223602} (\bibinfo {year} {2022})}\BibitemShut {NoStop}%
\bibitem [{\citenamefont {Vadiraj}\ \emph {et~al.}(2021)\citenamefont
  {Vadiraj}, \citenamefont {Ask}, \citenamefont {McConkey}, \citenamefont
  {Nsanzineza}, \citenamefont {Chang}, \citenamefont {Kockum},\ and\
  \citenamefont {Wilson}}]{Vadiraj2021}%
  \BibitemOpen
  \bibfield  {author} {\bibinfo {author} {\bibfnamefont {A.~M.}\ \bibnamefont
  {Vadiraj}}, \bibinfo {author} {\bibfnamefont {A.}~\bibnamefont {Ask}},
  \bibinfo {author} {\bibfnamefont {T.~G.}\ \bibnamefont {McConkey}}, \bibinfo
  {author} {\bibfnamefont {I.}~\bibnamefont {Nsanzineza}}, \bibinfo {author}
  {\bibfnamefont {C.~W.~S.}\ \bibnamefont {Chang}}, \bibinfo {author}
  {\bibfnamefont {A.~F.}\ \bibnamefont {Kockum}},\ and\ \bibinfo {author}
  {\bibfnamefont {C.~M.}\ \bibnamefont {Wilson}},\ }\bibfield  {title}
  {\bibinfo {title} {Engineering the level structure of a giant artificial atom
  in waveguide quantum electrodynamics},\ }\href
  {https://doi.org/10.1103/PhysRevA.103.023710} {\bibfield  {journal} {\bibinfo
   {journal} {Phys. Rev. A}\ }\textbf {\bibinfo {volume} {103}},\ \bibinfo
  {pages} {023710} (\bibinfo {year} {2021})}\BibitemShut {NoStop}%
\bibitem [{\citenamefont {Kannan}\ \emph {et~al.}(2020)\citenamefont {Kannan},
  \citenamefont {Ruckriegel}, \citenamefont {Campbell}, \citenamefont
  {Frisk~Kockum}, \citenamefont {Braum{\"u}ller}, \citenamefont {Kim},
  \citenamefont {Kjaergaard}, \citenamefont {Krantz}, \citenamefont {Melville},
  \citenamefont {Niedzielski} \emph {et~al.}}]{Kannan2020}%
  \BibitemOpen
  \bibfield  {author} {\bibinfo {author} {\bibfnamefont {B.}~\bibnamefont
  {Kannan}}, \bibinfo {author} {\bibfnamefont {M.~J.}\ \bibnamefont
  {Ruckriegel}}, \bibinfo {author} {\bibfnamefont {D.~L.}\ \bibnamefont
  {Campbell}}, \bibinfo {author} {\bibfnamefont {A.}~\bibnamefont
  {Frisk~Kockum}}, \bibinfo {author} {\bibfnamefont {J.}~\bibnamefont
  {Braum{\"u}ller}}, \bibinfo {author} {\bibfnamefont {D.~K.}\ \bibnamefont
  {Kim}}, \bibinfo {author} {\bibfnamefont {M.}~\bibnamefont {Kjaergaard}},
  \bibinfo {author} {\bibfnamefont {P.}~\bibnamefont {Krantz}}, \bibinfo
  {author} {\bibfnamefont {A.}~\bibnamefont {Melville}}, \bibinfo {author}
  {\bibfnamefont {B.~M.}\ \bibnamefont {Niedzielski}}, \emph {et~al.},\
  }\bibfield  {title} {\bibinfo {title} {Waveguide quantum electrodynamics with
  superconducting artificial giant atoms},\ }\href
  {https://doi.org/https://doi.org/10.1038/s41586-020-2529-9} {\bibfield
  {journal} {\bibinfo  {journal} {Nature}\ }\textbf {\bibinfo {volume} {583}},\
  \bibinfo {pages} {775} (\bibinfo {year} {2020})}\BibitemShut {NoStop}%
\bibitem [{\citenamefont {Manenti}\ \emph {et~al.}(2017)\citenamefont
  {Manenti}, \citenamefont {Kockum}, \citenamefont {Patterson}, \citenamefont
  {Behrle}, \citenamefont {Rahamim}, \citenamefont {Tancredi}, \citenamefont
  {Nori},\ and\ \citenamefont {Leek}}]{Manenti2017}%
  \BibitemOpen
  \bibfield  {author} {\bibinfo {author} {\bibfnamefont {R.}~\bibnamefont
  {Manenti}}, \bibinfo {author} {\bibfnamefont {A.~F.}\ \bibnamefont {Kockum}},
  \bibinfo {author} {\bibfnamefont {A.}~\bibnamefont {Patterson}}, \bibinfo
  {author} {\bibfnamefont {T.}~\bibnamefont {Behrle}}, \bibinfo {author}
  {\bibfnamefont {J.}~\bibnamefont {Rahamim}}, \bibinfo {author} {\bibfnamefont
  {G.}~\bibnamefont {Tancredi}}, \bibinfo {author} {\bibfnamefont
  {F.}~\bibnamefont {Nori}},\ and\ \bibinfo {author} {\bibfnamefont {P.~J.}\
  \bibnamefont {Leek}},\ }\bibfield  {title} {\bibinfo {title} {Circuit quantum
  acoustodynamics with surface acoustic waves},\ }\href
  {https://doi.org/https://doi.org/10.1038/s41467-017-01063-9} {\bibfield
  {journal} {\bibinfo  {journal} {Nature communications}\ }\textbf {\bibinfo
  {volume} {8}},\ \bibinfo {pages} {975} (\bibinfo {year} {2017})}\BibitemShut
  {NoStop}%
\bibitem [{\citenamefont {Andersson}\ \emph {et~al.}(2019)\citenamefont
  {Andersson}, \citenamefont {Suri}, \citenamefont {Guo}, \citenamefont
  {Aref},\ and\ \citenamefont {Delsing}}]{Andersson2019}%
  \BibitemOpen
  \bibfield  {author} {\bibinfo {author} {\bibfnamefont {G.}~\bibnamefont
  {Andersson}}, \bibinfo {author} {\bibfnamefont {B.}~\bibnamefont {Suri}},
  \bibinfo {author} {\bibfnamefont {L.}~\bibnamefont {Guo}}, \bibinfo {author}
  {\bibfnamefont {T.}~\bibnamefont {Aref}},\ and\ \bibinfo {author}
  {\bibfnamefont {P.}~\bibnamefont {Delsing}},\ }\bibfield  {title} {\bibinfo
  {title} {Non-exponential decay of a giant artificial atom},\ }\href
  {https://doi.org/10.1038/s41567-019-0605-6} {\bibfield  {journal} {\bibinfo
  {journal} {Nature Physics}\ }\textbf {\bibinfo {volume} {15}},\ \bibinfo
  {pages} {1123} (\bibinfo {year} {2019})}\BibitemShut {NoStop}%
\bibitem [{\citenamefont {Frisk~Kockum}\ \emph {et~al.}(2014)\citenamefont
  {Frisk~Kockum}, \citenamefont {Delsing},\ and\ \citenamefont
  {Johansson}}]{Frisk2014}%
  \BibitemOpen
  \bibfield  {author} {\bibinfo {author} {\bibfnamefont {A.}~\bibnamefont
  {Frisk~Kockum}}, \bibinfo {author} {\bibfnamefont {P.}~\bibnamefont
  {Delsing}},\ and\ \bibinfo {author} {\bibfnamefont {G.}~\bibnamefont
  {Johansson}},\ }\bibfield  {title} {\bibinfo {title} {Designing
  frequency-dependent relaxation rates and lamb shifts for a giant artificial
  atom},\ }\href {https://doi.org/10.1103/PhysRevA.90.013837} {\bibfield
  {journal} {\bibinfo  {journal} {Phys. Rev. A}\ }\textbf {\bibinfo {volume}
  {90}},\ \bibinfo {pages} {013837} (\bibinfo {year} {2014})}\BibitemShut
  {NoStop}%
\bibitem [{\citenamefont {Guo}\ \emph {et~al.}(2020{\natexlab{a}})\citenamefont
  {Guo}, \citenamefont {Wang}, \citenamefont {Purdy},\ and\ \citenamefont
  {Taylor}}]{Guo2020_1}%
  \BibitemOpen
  \bibfield  {author} {\bibinfo {author} {\bibfnamefont {S.}~\bibnamefont
  {Guo}}, \bibinfo {author} {\bibfnamefont {Y.}~\bibnamefont {Wang}}, \bibinfo
  {author} {\bibfnamefont {T.}~\bibnamefont {Purdy}},\ and\ \bibinfo {author}
  {\bibfnamefont {J.}~\bibnamefont {Taylor}},\ }\bibfield  {title} {\bibinfo
  {title} {Beyond spontaneous emission: Giant atom bounded in the continuum},\
  }\href {https://doi.org/10.1103/PhysRevA.102.033706} {\bibfield  {journal}
  {\bibinfo  {journal} {Phys. Rev. A}\ }\textbf {\bibinfo {volume} {102}},\
  \bibinfo {pages} {033706} (\bibinfo {year} {2020}{\natexlab{a}})}\BibitemShut
  {NoStop}%
\bibitem [{\citenamefont {Guo}\ \emph {et~al.}(2020{\natexlab{b}})\citenamefont
  {Guo}, \citenamefont {Kockum}, \citenamefont {Marquardt},\ and\ \citenamefont
  {Johansson}}]{Guo2020_2}%
  \BibitemOpen
  \bibfield  {author} {\bibinfo {author} {\bibfnamefont {L.}~\bibnamefont
  {Guo}}, \bibinfo {author} {\bibfnamefont {A.~F.}\ \bibnamefont {Kockum}},
  \bibinfo {author} {\bibfnamefont {F.}~\bibnamefont {Marquardt}},\ and\
  \bibinfo {author} {\bibfnamefont {G.}~\bibnamefont {Johansson}},\ }\bibfield
  {title} {\bibinfo {title} {Oscillating bound states for a giant atom},\
  }\href {https://doi.org/10.1103/PhysRevResearch.2.043014} {\bibfield
  {journal} {\bibinfo  {journal} {Phys. Rev. Res.}\ }\textbf {\bibinfo {volume}
  {2}},\ \bibinfo {pages} {043014} (\bibinfo {year}
  {2020}{\natexlab{b}})}\BibitemShut {NoStop}%
\bibitem [{\citenamefont {Zhao}\ and\ \citenamefont {Wang}(2020)}]{Zhao2020}%
  \BibitemOpen
  \bibfield  {author} {\bibinfo {author} {\bibfnamefont {W.}~\bibnamefont
  {Zhao}}\ and\ \bibinfo {author} {\bibfnamefont {Z.}~\bibnamefont {Wang}},\
  }\bibfield  {title} {\bibinfo {title} {Single-photon scattering and bound
  states in an atom-waveguide system with two or multiple coupling points},\
  }\href {https://doi.org/10.1103/PhysRevA.101.053855} {\bibfield  {journal}
  {\bibinfo  {journal} {Phys. Rev. A}\ }\textbf {\bibinfo {volume} {101}},\
  \bibinfo {pages} {053855} (\bibinfo {year} {2020})}\BibitemShut {NoStop}%
\bibitem [{\citenamefont {Noachtar}\ \emph {et~al.}(2022)\citenamefont
  {Noachtar}, \citenamefont {Kn\"orzer},\ and\ \citenamefont
  {Jonsson}}]{Noachtar2022}%
  \BibitemOpen
  \bibfield  {author} {\bibinfo {author} {\bibfnamefont {D.~D.}\ \bibnamefont
  {Noachtar}}, \bibinfo {author} {\bibfnamefont {J.}~\bibnamefont
  {Kn\"orzer}},\ and\ \bibinfo {author} {\bibfnamefont {R.~H.}\ \bibnamefont
  {Jonsson}},\ }\bibfield  {title} {\bibinfo {title} {Nonperturbative treatment
  of giant atoms using chain transformations},\ }\href
  {https://doi.org/10.1103/PhysRevA.106.013702} {\bibfield  {journal} {\bibinfo
   {journal} {Phys. Rev. A}\ }\textbf {\bibinfo {volume} {106}},\ \bibinfo
  {pages} {013702} (\bibinfo {year} {2022})}\BibitemShut {NoStop}%
\bibitem [{\citenamefont {Zhou}\ \emph {et~al.}(2023)\citenamefont {Zhou},
  \citenamefont {Yin},\ and\ \citenamefont {Liao}}]{Zhou2023}%
  \BibitemOpen
  \bibfield  {author} {\bibinfo {author} {\bibfnamefont {J.}~\bibnamefont
  {Zhou}}, \bibinfo {author} {\bibfnamefont {X.-L.}\ \bibnamefont {Yin}},\ and\
  \bibinfo {author} {\bibfnamefont {J.-Q.}\ \bibnamefont {Liao}},\ }\bibfield
  {title} {\bibinfo {title} {Chiral and nonreciprocal single-photon scattering
  in a chiral-giant-molecule waveguide-qed system},\ }\href
  {https://doi.org/10.1103/PhysRevA.107.063703} {\bibfield  {journal} {\bibinfo
   {journal} {Phys. Rev. A}\ }\textbf {\bibinfo {volume} {107}},\ \bibinfo
  {pages} {063703} (\bibinfo {year} {2023})}\BibitemShut {NoStop}%
\bibitem [{\citenamefont {Kockum}\ \emph {et~al.}(2018)\citenamefont {Kockum},
  \citenamefont {Johansson},\ and\ \citenamefont {Nori}}]{Kockum2018}%
  \BibitemOpen
  \bibfield  {author} {\bibinfo {author} {\bibfnamefont {A.~F.}\ \bibnamefont
  {Kockum}}, \bibinfo {author} {\bibfnamefont {G.}~\bibnamefont {Johansson}},\
  and\ \bibinfo {author} {\bibfnamefont {F.}~\bibnamefont {Nori}},\ }\bibfield
  {title} {\bibinfo {title} {Decoherence-free interaction between giant atoms
  in waveguide quantum electrodynamics},\ }\href
  {https://doi.org/10.1103/PhysRevLett.120.140404} {\bibfield  {journal}
  {\bibinfo  {journal} {Phys. Rev. Lett.}\ }\textbf {\bibinfo {volume} {120}},\
  \bibinfo {pages} {140404} (\bibinfo {year} {2018})}\BibitemShut {NoStop}%
\bibitem [{\citenamefont {Du}\ \emph {et~al.}(2023)\citenamefont {Du},
  \citenamefont {Guo},\ and\ \citenamefont {Li}}]{Du2023}%
  \BibitemOpen
  \bibfield  {author} {\bibinfo {author} {\bibfnamefont {L.}~\bibnamefont
  {Du}}, \bibinfo {author} {\bibfnamefont {L.}~\bibnamefont {Guo}},\ and\
  \bibinfo {author} {\bibfnamefont {Y.}~\bibnamefont {Li}},\ }\bibfield
  {title} {\bibinfo {title} {Complex decoherence-free interactions between
  giant atoms},\ }\href {https://doi.org/10.1103/PhysRevA.107.023705}
  {\bibfield  {journal} {\bibinfo  {journal} {Phys. Rev. A}\ }\textbf {\bibinfo
  {volume} {107}},\ \bibinfo {pages} {023705} (\bibinfo {year}
  {2023})}\BibitemShut {NoStop}%
\bibitem [{\citenamefont {Yin}\ \emph {et~al.}(2022)\citenamefont {Yin},
  \citenamefont {Luo},\ and\ \citenamefont {Liao}}]{Yin2022}%
  \BibitemOpen
  \bibfield  {author} {\bibinfo {author} {\bibfnamefont {X.-L.}\ \bibnamefont
  {Yin}}, \bibinfo {author} {\bibfnamefont {W.-B.}\ \bibnamefont {Luo}},\ and\
  \bibinfo {author} {\bibfnamefont {J.-Q.}\ \bibnamefont {Liao}},\ }\bibfield
  {title} {\bibinfo {title} {Non-markovian disentanglement dynamics in
  double-giant-atom waveguide-qed systems},\ }\href
  {https://doi.org/10.1103/PhysRevA.106.063703} {\bibfield  {journal} {\bibinfo
   {journal} {Phys. Rev. A}\ }\textbf {\bibinfo {volume} {106}},\ \bibinfo
  {pages} {063703} (\bibinfo {year} {2022})}\BibitemShut {NoStop}%
\bibitem [{\citenamefont {Yin}\ and\ \citenamefont {Liao}(2023)}]{Yin2023}%
  \BibitemOpen
  \bibfield  {author} {\bibinfo {author} {\bibfnamefont {X.-L.}\ \bibnamefont
  {Yin}}\ and\ \bibinfo {author} {\bibfnamefont {J.-Q.}\ \bibnamefont {Liao}},\
  }\bibfield  {title} {\bibinfo {title} {Generation of two-giant-atom
  entanglement in waveguide-qed systems},\ }\href
  {https://doi.org/10.1103/PhysRevA.108.023728} {\bibfield  {journal} {\bibinfo
   {journal} {Phys. Rev. A}\ }\textbf {\bibinfo {volume} {108}},\ \bibinfo
  {pages} {023728} (\bibinfo {year} {2023})}\BibitemShut {NoStop}%
\bibitem [{\citenamefont {Yu}\ \emph {et~al.}(2021)\citenamefont {Yu},
  \citenamefont {Wang},\ and\ \citenamefont {Wu}}]{Yu2021}%
  \BibitemOpen
  \bibfield  {author} {\bibinfo {author} {\bibfnamefont {H.}~\bibnamefont
  {Yu}}, \bibinfo {author} {\bibfnamefont {Z.}~\bibnamefont {Wang}},\ and\
  \bibinfo {author} {\bibfnamefont {J.-H.}\ \bibnamefont {Wu}},\ }\bibfield
  {title} {\bibinfo {title} {Entanglement preparation and nonreciprocal
  excitation evolution in giant atoms by controllable dissipation and
  coupling},\ }\href {https://doi.org/10.1103/PhysRevA.104.013720} {\bibfield
  {journal} {\bibinfo  {journal} {Phys. Rev. A}\ }\textbf {\bibinfo {volume}
  {104}},\ \bibinfo {pages} {013720} (\bibinfo {year} {2021})}\BibitemShut
  {NoStop}%
\bibitem [{\citenamefont {Santos}\ and\ \citenamefont
  {Bachelard}(2023)}]{Santos2023}%
  \BibitemOpen
  \bibfield  {author} {\bibinfo {author} {\bibfnamefont {A.~C.}\ \bibnamefont
  {Santos}}\ and\ \bibinfo {author} {\bibfnamefont {R.}~\bibnamefont
  {Bachelard}},\ }\bibfield  {title} {\bibinfo {title} {Generation of maximally
  entangled long-lived states with giant atoms in a waveguide},\ }\href
  {https://doi.org/10.1103/PhysRevLett.130.053601} {\bibfield  {journal}
  {\bibinfo  {journal} {Phys. Rev. Lett.}\ }\textbf {\bibinfo {volume} {130}},\
  \bibinfo {pages} {053601} (\bibinfo {year} {2023})}\BibitemShut {NoStop}%
\bibitem [{\citenamefont {Chen}\ and\ \citenamefont
  {Frisk~Kockum}(2025)}]{Chen2025}%
  \BibitemOpen
  \bibfield  {author} {\bibinfo {author} {\bibfnamefont {G.}~\bibnamefont
  {Chen}}\ and\ \bibinfo {author} {\bibfnamefont {A.}~\bibnamefont
  {Frisk~Kockum}},\ }\bibfield  {title} {\bibinfo {title} {Simulating open
  quantum systems with giant atoms},\ }\href
  {https://doi.org/10.1088/2058-9565/adb2bd} {\bibfield  {journal} {\bibinfo
  {journal} {Quantum Science and Technology}\ }\textbf {\bibinfo {volume}
  {10}},\ \bibinfo {pages} {025028} (\bibinfo {year} {2025})}\BibitemShut
  {NoStop}%
\bibitem [{\citenamefont {Du}\ \emph {et~al.}(2025)\citenamefont {Du},
  \citenamefont {Wang}, \citenamefont {Kockum},\ and\ \citenamefont
  {Splettstoesser}}]{Du2025}%
  \BibitemOpen
  \bibfield  {author} {\bibinfo {author} {\bibfnamefont {L.}~\bibnamefont
  {Du}}, \bibinfo {author} {\bibfnamefont {X.}~\bibnamefont {Wang}}, \bibinfo
  {author} {\bibfnamefont {A.~F.}\ \bibnamefont {Kockum}},\ and\ \bibinfo
  {author} {\bibfnamefont {J.}~\bibnamefont {Splettstoesser}},\ }\bibfield
  {title} {\bibinfo {title} {Dressed interference in giant superatoms:
  Entanglement generation and transfer},\ }\href
  {https://doi.org/10.1103/crzs-k718} {\bibfield  {journal} {\bibinfo
  {journal} {Phys. Rev. Lett.}\ }\textbf {\bibinfo {volume} {135}},\ \bibinfo
  {pages} {223601} (\bibinfo {year} {2025})}\BibitemShut {NoStop}%
\bibitem [{\citenamefont {Grimsmo}(2015)}]{Grimsmo2015}%
  \BibitemOpen
  \bibfield  {author} {\bibinfo {author} {\bibfnamefont {A.~L.}\ \bibnamefont
  {Grimsmo}},\ }\bibfield  {title} {\bibinfo {title} {Time-delayed quantum
  feedback control},\ }\href {https://doi.org/10.1103/PhysRevLett.115.060402}
  {\bibfield  {journal} {\bibinfo  {journal} {Phys. Rev. Lett.}\ }\textbf
  {\bibinfo {volume} {115}},\ \bibinfo {pages} {060402} (\bibinfo {year}
  {2015})}\BibitemShut {NoStop}%
\bibitem [{\citenamefont {Pichler}\ and\ \citenamefont
  {Zoller}(2016)}]{Pichler2016}%
  \BibitemOpen
  \bibfield  {author} {\bibinfo {author} {\bibfnamefont {H.}~\bibnamefont
  {Pichler}}\ and\ \bibinfo {author} {\bibfnamefont {P.}~\bibnamefont
  {Zoller}},\ }\bibfield  {title} {\bibinfo {title} {Photonic circuits with
  time delays and quantum feedback},\ }\href
  {https://doi.org/10.1103/PhysRevLett.116.093601} {\bibfield  {journal}
  {\bibinfo  {journal} {Phys. Rev. Lett.}\ }\textbf {\bibinfo {volume} {116}},\
  \bibinfo {pages} {093601} (\bibinfo {year} {2016})}\BibitemShut {NoStop}%
\bibitem [{\citenamefont {Caruso}\ \emph {et~al.}(2014)\citenamefont {Caruso},
  \citenamefont {Giovannetti}, \citenamefont {Lupo},\ and\ \citenamefont
  {Mancini}}]{Caruso2014}%
  \BibitemOpen
  \bibfield  {author} {\bibinfo {author} {\bibfnamefont {F.}~\bibnamefont
  {Caruso}}, \bibinfo {author} {\bibfnamefont {V.}~\bibnamefont {Giovannetti}},
  \bibinfo {author} {\bibfnamefont {C.}~\bibnamefont {Lupo}},\ and\ \bibinfo
  {author} {\bibfnamefont {S.}~\bibnamefont {Mancini}},\ }\bibfield  {title}
  {\bibinfo {title} {Quantum channels and memory effects},\ }\href
  {https://doi.org/10.1103/RevModPhys.86.1203} {\bibfield  {journal} {\bibinfo
  {journal} {Rev. Mod. Phys.}\ }\textbf {\bibinfo {volume} {86}},\ \bibinfo
  {pages} {1203} (\bibinfo {year} {2014})}\BibitemShut {NoStop}%
\bibitem [{\citenamefont {Breuer}\ \emph {et~al.}(2016)\citenamefont {Breuer},
  \citenamefont {Laine}, \citenamefont {Piilo},\ and\ \citenamefont
  {Vacchini}}]{Breuer2016}%
  \BibitemOpen
  \bibfield  {author} {\bibinfo {author} {\bibfnamefont {H.-P.}\ \bibnamefont
  {Breuer}}, \bibinfo {author} {\bibfnamefont {E.-M.}\ \bibnamefont {Laine}},
  \bibinfo {author} {\bibfnamefont {J.}~\bibnamefont {Piilo}},\ and\ \bibinfo
  {author} {\bibfnamefont {B.}~\bibnamefont {Vacchini}},\ }\bibfield  {title}
  {\bibinfo {title} {Colloquium: Non-markovian dynamics in open quantum
  systems},\ }\href {https://doi.org/10.1103/RevModPhys.88.021002} {\bibfield
  {journal} {\bibinfo  {journal} {Rev. Mod. Phys.}\ }\textbf {\bibinfo {volume}
  {88}},\ \bibinfo {pages} {021002} (\bibinfo {year} {2016})}\BibitemShut
  {NoStop}%
\bibitem [{\citenamefont {de~Vega}\ and\ \citenamefont
  {Alonso}(2017)}]{deVega2017}%
  \BibitemOpen
  \bibfield  {author} {\bibinfo {author} {\bibfnamefont {I.}~\bibnamefont
  {de~Vega}}\ and\ \bibinfo {author} {\bibfnamefont {D.}~\bibnamefont
  {Alonso}},\ }\bibfield  {title} {\bibinfo {title} {Dynamics of non-markovian
  open quantum systems},\ }\href {https://doi.org/10.1103/RevModPhys.89.015001}
  {\bibfield  {journal} {\bibinfo  {journal} {Rev. Mod. Phys.}\ }\textbf
  {\bibinfo {volume} {89}},\ \bibinfo {pages} {015001} (\bibinfo {year}
  {2017})}\BibitemShut {NoStop}%
\bibitem [{\citenamefont {Yu}\ \emph {et~al.}(2025)\citenamefont {Yu},
  \citenamefont {Ohst}, \citenamefont {Nguyen},\ and\ \citenamefont
  {Nimmrichter}}]{Yu2025}%
  \BibitemOpen
  \bibfield  {author} {\bibinfo {author} {\bibfnamefont {M.}~\bibnamefont
  {Yu}}, \bibinfo {author} {\bibfnamefont {T.-A.}\ \bibnamefont {Ohst}},
  \bibinfo {author} {\bibfnamefont {H.-C.}\ \bibnamefont {Nguyen}},\ and\
  \bibinfo {author} {\bibfnamefont {S.}~\bibnamefont {Nimmrichter}},\ }\href
  {https://arxiv.org/abs/2504.08605} {\bibinfo {title} {Quantum memory in
  spontaneous emission processes}} (\bibinfo {year} {2025}),\ \Eprint
  {https://arxiv.org/abs/2504.08605} {arXiv:2504.08605 [quant-ph]} \BibitemShut
  {NoStop}%
\bibitem [{\citenamefont {Wang}\ \emph {et~al.}(2017)\citenamefont {Wang},
  \citenamefont {Um}, \citenamefont {Zhang}, \citenamefont {An}, \citenamefont
  {Lyu}, \citenamefont {Zhang}, \citenamefont {Duan}, \citenamefont {Yum},\
  and\ \citenamefont {Kim}}]{Wang2017}%
  \BibitemOpen
  \bibfield  {author} {\bibinfo {author} {\bibfnamefont {Y.}~\bibnamefont
  {Wang}}, \bibinfo {author} {\bibfnamefont {M.}~\bibnamefont {Um}}, \bibinfo
  {author} {\bibfnamefont {J.}~\bibnamefont {Zhang}}, \bibinfo {author}
  {\bibfnamefont {S.}~\bibnamefont {An}}, \bibinfo {author} {\bibfnamefont
  {M.}~\bibnamefont {Lyu}}, \bibinfo {author} {\bibfnamefont {J.-N.}\
  \bibnamefont {Zhang}}, \bibinfo {author} {\bibfnamefont {L.-M.}\ \bibnamefont
  {Duan}}, \bibinfo {author} {\bibfnamefont {D.}~\bibnamefont {Yum}},\ and\
  \bibinfo {author} {\bibfnamefont {K.}~\bibnamefont {Kim}},\ }\bibfield
  {title} {\bibinfo {title} {Single-qubit quantum memory exceeding ten-minute
  coherence time},\ }\href@noop {} {\bibfield  {journal} {\bibinfo  {journal}
  {Nature Photonics}\ }\textbf {\bibinfo {volume} {11}},\ \bibinfo {pages}
  {646} (\bibinfo {year} {2017})}\BibitemShut {NoStop}%
\bibitem [{\citenamefont {Binder}\ \emph {et~al.}(2019)\citenamefont {Binder},
  \citenamefont {Correa}, \citenamefont {Gogolin}, \citenamefont {Anders},\
  and\ \citenamefont {Adesso}}]{Binder2019}%
  \BibitemOpen
  \bibfield  {author} {\bibinfo {author} {\bibfnamefont {F.}~\bibnamefont
  {Binder}}, \bibinfo {author} {\bibfnamefont {L.}~\bibnamefont {Correa}},
  \bibinfo {author} {\bibfnamefont {C.}~\bibnamefont {Gogolin}}, \bibinfo
  {author} {\bibfnamefont {J.}~\bibnamefont {Anders}},\ and\ \bibinfo {author}
  {\bibfnamefont {G.}~\bibnamefont {Adesso}},\ }\href
  {https://books.google.de/books?id=IQlouQEACAAJ} {\emph {\bibinfo {title}
  {Thermodynamics in the Quantum Regime: Fundamental Aspects and New
  Directions}}},\ Fundamental Theories of Physics\ (\bibinfo  {publisher}
  {Springer International Publishing},\ \bibinfo {year} {2019})\BibitemShut
  {NoStop}%
\bibitem [{\citenamefont {Newman}\ \emph {et~al.}(2017)\citenamefont {Newman},
  \citenamefont {Mintert},\ and\ \citenamefont {Nazir}}]{Newman2017}%
  \BibitemOpen
  \bibfield  {author} {\bibinfo {author} {\bibfnamefont {D.}~\bibnamefont
  {Newman}}, \bibinfo {author} {\bibfnamefont {F.}~\bibnamefont {Mintert}},\
  and\ \bibinfo {author} {\bibfnamefont {A.}~\bibnamefont {Nazir}},\ }\bibfield
   {title} {\bibinfo {title} {Performance of a quantum heat engine at strong
  reservoir coupling},\ }\href {https://doi.org/10.1103/PhysRevE.95.032139}
  {\bibfield  {journal} {\bibinfo  {journal} {Phys. Rev. E}\ }\textbf {\bibinfo
  {volume} {95}},\ \bibinfo {pages} {032139} (\bibinfo {year}
  {2017})}\BibitemShut {NoStop}%
\bibitem [{\citenamefont {Valentini}(1991)}]{Valentin1991}%
  \BibitemOpen
  \bibfield  {author} {\bibinfo {author} {\bibfnamefont {A.}~\bibnamefont
  {Valentini}},\ }\bibfield  {title} {\bibinfo {title} {Non-local correlations
  in quantum electrodynamics},\ }\href
  {https://doi.org/https://doi.org/10.1016/0375-9601(91)90952-5} {\bibfield
  {journal} {\bibinfo  {journal} {Physics Letters A}\ }\textbf {\bibinfo
  {volume} {153}},\ \bibinfo {pages} {321} (\bibinfo {year}
  {1991})}\BibitemShut {NoStop}%
\bibitem [{\citenamefont {Le\'on}\ and\ \citenamefont
  {Sab\'{\i}n}(2009)}]{Juan2009}%
  \BibitemOpen
  \bibfield  {author} {\bibinfo {author} {\bibfnamefont {J.}~\bibnamefont
  {Le\'on}}\ and\ \bibinfo {author} {\bibfnamefont {C.}~\bibnamefont
  {Sab\'{\i}n}},\ }\bibfield  {title} {\bibinfo {title} {Generation of
  atom-atom correlations inside and outside the mutual light cone},\ }\href
  {https://doi.org/10.1103/PhysRevA.79.012304} {\bibfield  {journal} {\bibinfo
  {journal} {Phys. Rev. A}\ }\textbf {\bibinfo {volume} {79}},\ \bibinfo
  {pages} {012304} (\bibinfo {year} {2009})}\BibitemShut {NoStop}%
\bibitem [{\citenamefont {Sab\'{\i}n}\ \emph {et~al.}(2010)\citenamefont
  {Sab\'{\i}n}, \citenamefont {Garc\'{\i}a-Ripoll}, \citenamefont {Solano},\
  and\ \citenamefont {Le\'on}}]{Sabin2010}%
  \BibitemOpen
  \bibfield  {author} {\bibinfo {author} {\bibfnamefont {C.}~\bibnamefont
  {Sab\'{\i}n}}, \bibinfo {author} {\bibfnamefont {J.~J.}\ \bibnamefont
  {Garc\'{\i}a-Ripoll}}, \bibinfo {author} {\bibfnamefont {E.}~\bibnamefont
  {Solano}},\ and\ \bibinfo {author} {\bibfnamefont {J.}~\bibnamefont
  {Le\'on}},\ }\bibfield  {title} {\bibinfo {title} {Dynamics of entanglement
  via propagating microwave photons},\ }\href
  {https://doi.org/10.1103/PhysRevB.81.184501} {\bibfield  {journal} {\bibinfo
  {journal} {Phys. Rev. B}\ }\textbf {\bibinfo {volume} {81}},\ \bibinfo
  {pages} {184501} (\bibinfo {year} {2010})}\BibitemShut {NoStop}%
\bibitem [{\citenamefont {Guo}\ \emph {et~al.}(2017)\citenamefont {Guo},
  \citenamefont {Grimsmo}, \citenamefont {Kockum}, \citenamefont {Pletyukhov},\
  and\ \citenamefont {Johansson}}]{Guo2017}%
  \BibitemOpen
  \bibfield  {author} {\bibinfo {author} {\bibfnamefont {L.}~\bibnamefont
  {Guo}}, \bibinfo {author} {\bibfnamefont {A.}~\bibnamefont {Grimsmo}},
  \bibinfo {author} {\bibfnamefont {A.~F.}\ \bibnamefont {Kockum}}, \bibinfo
  {author} {\bibfnamefont {M.}~\bibnamefont {Pletyukhov}},\ and\ \bibinfo
  {author} {\bibfnamefont {G.}~\bibnamefont {Johansson}},\ }\bibfield  {title}
  {\bibinfo {title} {Giant acoustic atom: A single quantum system with a
  deterministic time delay},\ }\href
  {https://doi.org/10.1103/PhysRevA.95.053821} {\bibfield  {journal} {\bibinfo
  {journal} {Phys. Rev. A}\ }\textbf {\bibinfo {volume} {95}},\ \bibinfo
  {pages} {053821} (\bibinfo {year} {2017})}\BibitemShut {NoStop}%
\bibitem [{\citenamefont {Magnifico}\ \emph {et~al.}(2025)\citenamefont
  {Magnifico}, \citenamefont {Maffei}, \citenamefont {Pomarico}, \citenamefont
  {Das}, \citenamefont {Facchi}, \citenamefont {Pascazio},\ and\ \citenamefont
  {Pepe}}]{Magnifico2025}%
  \BibitemOpen
  \bibfield  {author} {\bibinfo {author} {\bibfnamefont {G.}~\bibnamefont
  {Magnifico}}, \bibinfo {author} {\bibfnamefont {M.}~\bibnamefont {Maffei}},
  \bibinfo {author} {\bibfnamefont {D.}~\bibnamefont {Pomarico}}, \bibinfo
  {author} {\bibfnamefont {D.}~\bibnamefont {Das}}, \bibinfo {author}
  {\bibfnamefont {P.}~\bibnamefont {Facchi}}, \bibinfo {author} {\bibfnamefont
  {S.}~\bibnamefont {Pascazio}},\ and\ \bibinfo {author} {\bibfnamefont
  {F.~V.}\ \bibnamefont {Pepe}},\ }\bibfield  {title} {\bibinfo {title}
  {Non-markovian dynamics of generation of bound states in the continuum via
  single-photon scattering},\ }\href {https://doi.org/10.1103/gtf6-zb57}
  {\bibfield  {journal} {\bibinfo  {journal} {Phys. Rev. Res.}\ }\textbf
  {\bibinfo {volume} {7}},\ \bibinfo {pages} {033249} (\bibinfo {year}
  {2025})}\BibitemShut {NoStop}%
\bibitem [{\citenamefont {Chin}\ \emph {et~al.}(2010)\citenamefont {Chin},
  \citenamefont {Rivas}, \citenamefont {Huelga},\ and\ \citenamefont
  {Plenio}}]{Chin2010}%
  \BibitemOpen
  \bibfield  {author} {\bibinfo {author} {\bibfnamefont {A.~W.}\ \bibnamefont
  {Chin}}, \bibinfo {author} {\bibfnamefont {A.}~\bibnamefont {Rivas}},
  \bibinfo {author} {\bibfnamefont {S.~F.}\ \bibnamefont {Huelga}},\ and\
  \bibinfo {author} {\bibfnamefont {M.~B.}\ \bibnamefont {Plenio}},\ }\bibfield
   {title} {\bibinfo {title} {Exact mapping between system-reservoir quantum
  models and semi-infinite discrete chains using orthogonal polynomials},\
  }\href {https://doi.org/10.1063/1.3490188} {\bibfield  {journal} {\bibinfo
  {journal} {Journal of Mathematical Physics}\ }\textbf {\bibinfo {volume}
  {51}},\ \bibinfo {pages} {092109} (\bibinfo {year} {2010})}\BibitemShut
  {NoStop}%
\bibitem [{\citenamefont {Strathearn}\ \emph {et~al.}(2018)\citenamefont
  {Strathearn}, \citenamefont {Kirton}, \citenamefont {Kilda}, \citenamefont
  {Keeling},\ and\ \citenamefont {Lovett}}]{Strathearn2018}%
  \BibitemOpen
  \bibfield  {author} {\bibinfo {author} {\bibfnamefont {A.}~\bibnamefont
  {Strathearn}}, \bibinfo {author} {\bibfnamefont {P.}~\bibnamefont {Kirton}},
  \bibinfo {author} {\bibfnamefont {D.}~\bibnamefont {Kilda}}, \bibinfo
  {author} {\bibfnamefont {J.}~\bibnamefont {Keeling}},\ and\ \bibinfo {author}
  {\bibfnamefont {B.~W.}\ \bibnamefont {Lovett}},\ }\bibfield  {title}
  {\bibinfo {title} {Efficient non-markovian quantum dynamics using
  time-evolving matrix product operators},\ }\href
  {https://doi.org/https://doi.org/10.1038/s41467-018-05617-3} {\bibfield
  {journal} {\bibinfo  {journal} {Nature communications}\ }\textbf {\bibinfo
  {volume} {9}},\ \bibinfo {pages} {3322} (\bibinfo {year} {2018})}\BibitemShut
  {NoStop}%
\bibitem [{\citenamefont {Link}\ \emph {et~al.}(2024)\citenamefont {Link},
  \citenamefont {Tu},\ and\ \citenamefont {Strunz}}]{Link2024}%
  \BibitemOpen
  \bibfield  {author} {\bibinfo {author} {\bibfnamefont {V.}~\bibnamefont
  {Link}}, \bibinfo {author} {\bibfnamefont {H.-H.}\ \bibnamefont {Tu}},\ and\
  \bibinfo {author} {\bibfnamefont {W.~T.}\ \bibnamefont {Strunz}},\ }\bibfield
   {title} {\bibinfo {title} {Open quantum system dynamics from infinite tensor
  network contraction},\ }\href
  {https://doi.org/10.1103/PhysRevLett.132.200403} {\bibfield  {journal}
  {\bibinfo  {journal} {Phys. Rev. Lett.}\ }\textbf {\bibinfo {volume} {132}},\
  \bibinfo {pages} {200403} (\bibinfo {year} {2024})}\BibitemShut {NoStop}%
\bibitem [{\citenamefont {Suess}\ \emph {et~al.}(2014)\citenamefont {Suess},
  \citenamefont {Eisfeld},\ and\ \citenamefont {Strunz}}]{Suess2014}%
  \BibitemOpen
  \bibfield  {author} {\bibinfo {author} {\bibfnamefont {D.}~\bibnamefont
  {Suess}}, \bibinfo {author} {\bibfnamefont {A.}~\bibnamefont {Eisfeld}},\
  and\ \bibinfo {author} {\bibfnamefont {W.~T.}\ \bibnamefont {Strunz}},\
  }\bibfield  {title} {\bibinfo {title} {Hierarchy of stochastic pure states
  for open quantum system dynamics},\ }\href
  {https://doi.org/10.1103/PhysRevLett.113.150403} {\bibfield  {journal}
  {\bibinfo  {journal} {Phys. Rev. Lett.}\ }\textbf {\bibinfo {volume} {113}},\
  \bibinfo {pages} {150403} (\bibinfo {year} {2014})}\BibitemShut {NoStop}%
\bibitem [{\citenamefont {Tanimura}\ and\ \citenamefont
  {Kubo}(1989)}]{Tanimura1989}%
  \BibitemOpen
  \bibfield  {author} {\bibinfo {author} {\bibfnamefont {Y.}~\bibnamefont
  {Tanimura}}\ and\ \bibinfo {author} {\bibfnamefont {R.}~\bibnamefont
  {Kubo}},\ }\bibfield  {title} {\bibinfo {title} {Two-time correlation
  functions of a system coupled to a heat bath with a gaussian-markoffian
  interaction},\ }\href {https://doi.org/10.1143/JPSJ.58.1199} {\bibfield
  {journal} {\bibinfo  {journal} {Journal of the Physical Society of Japan}\
  }\textbf {\bibinfo {volume} {58}},\ \bibinfo {pages} {1199} (\bibinfo {year}
  {1989})}\BibitemShut {NoStop}%
\bibitem [{\citenamefont {Tanimura}(2020)}]{Tanimura2020}%
  \BibitemOpen
  \bibfield  {author} {\bibinfo {author} {\bibfnamefont {Y.}~\bibnamefont
  {Tanimura}},\ }\bibfield  {title} {\bibinfo {title} {Numerically “exact”
  approach to open quantum dynamics: The hierarchical equations of motion
  (heom)},\ }\href@noop {} {\bibfield  {journal} {\bibinfo  {journal} {The
  Journal of Chemical Physics}\ }\textbf {\bibinfo {volume} {153}},\ \bibinfo
  {pages} {020901} (\bibinfo {year} {2020})}\BibitemShut {NoStop}%
\bibitem [{\citenamefont {Lambert}\ \emph {et~al.}(2023)\citenamefont
  {Lambert}, \citenamefont {Raheja}, \citenamefont {Cross}, \citenamefont
  {Menczel}, \citenamefont {Ahmed}, \citenamefont {Pitchford}, \citenamefont
  {Burgarth},\ and\ \citenamefont {Nori}}]{Lambert2023}%
  \BibitemOpen
  \bibfield  {author} {\bibinfo {author} {\bibfnamefont {N.}~\bibnamefont
  {Lambert}}, \bibinfo {author} {\bibfnamefont {T.}~\bibnamefont {Raheja}},
  \bibinfo {author} {\bibfnamefont {S.}~\bibnamefont {Cross}}, \bibinfo
  {author} {\bibfnamefont {P.}~\bibnamefont {Menczel}}, \bibinfo {author}
  {\bibfnamefont {S.}~\bibnamefont {Ahmed}}, \bibinfo {author} {\bibfnamefont
  {A.}~\bibnamefont {Pitchford}}, \bibinfo {author} {\bibfnamefont
  {D.}~\bibnamefont {Burgarth}},\ and\ \bibinfo {author} {\bibfnamefont
  {F.}~\bibnamefont {Nori}},\ }\bibfield  {title} {\bibinfo {title}
  {Qutip-bofin: A bosonic and fermionic numerical
  hierarchical-equations-of-motion library with applications in
  light-harvesting, quantum control, and single-molecule electronics},\ }\href
  {https://doi.org/10.1103/PhysRevResearch.5.013181} {\bibfield  {journal}
  {\bibinfo  {journal} {Phys. Rev. Res.}\ }\textbf {\bibinfo {volume} {5}},\
  \bibinfo {pages} {013181} (\bibinfo {year} {2023})}\BibitemShut {NoStop}%
\bibitem [{\citenamefont {Roy}\ and\ \citenamefont {Kailath}(1989)}]{Roy1989}%
  \BibitemOpen
  \bibfield  {author} {\bibinfo {author} {\bibfnamefont {R.}~\bibnamefont
  {Roy}}\ and\ \bibinfo {author} {\bibfnamefont {T.}~\bibnamefont {Kailath}},\
  }\bibfield  {title} {\bibinfo {title} {Esprit-estimation of signal parameters
  via rotational invariance techniques},\ }\href
  {https://doi.org/10.1109/29.32276} {\bibfield  {journal} {\bibinfo  {journal}
  {IEEE Transactions on Acoustics, Speech, and Signal Processing}\ }\textbf
  {\bibinfo {volume} {37}},\ \bibinfo {pages} {984} (\bibinfo {year}
  {1989})}\BibitemShut {NoStop}%
\bibitem [{\citenamefont {Takahashi}\ \emph {et~al.}(2024)\citenamefont
  {Takahashi}, \citenamefont {Rudge}, \citenamefont {Kaspar}, \citenamefont
  {Thoss},\ and\ \citenamefont {Borrelli}}]{Takahashi2024}%
  \BibitemOpen
  \bibfield  {author} {\bibinfo {author} {\bibfnamefont {H.}~\bibnamefont
  {Takahashi}}, \bibinfo {author} {\bibfnamefont {S.}~\bibnamefont {Rudge}},
  \bibinfo {author} {\bibfnamefont {C.}~\bibnamefont {Kaspar}}, \bibinfo
  {author} {\bibfnamefont {M.}~\bibnamefont {Thoss}},\ and\ \bibinfo {author}
  {\bibfnamefont {R.}~\bibnamefont {Borrelli}},\ }\bibfield  {title} {\bibinfo
  {title} {High accuracy exponential decomposition of bath correlation
  functions for arbitrary and structured spectral densities: Emerging
  methodologies and new approaches},\ }\href
  {https://doi.org/10.1063/5.0209348} {\bibfield  {journal} {\bibinfo
  {journal} {The Journal of Chemical Physics}\ }\textbf {\bibinfo {volume}
  {160}},\ \bibinfo {pages} {204105} (\bibinfo {year} {2024})}\BibitemShut
  {NoStop}%
\bibitem [{\citenamefont {Guo}\ \emph {et~al.}(2020{\natexlab{c}})\citenamefont
  {Guo}, \citenamefont {Kockum}, \citenamefont {Marquardt},\ and\ \citenamefont
  {Johansson}}]{Guo2020}%
  \BibitemOpen
  \bibfield  {author} {\bibinfo {author} {\bibfnamefont {L.}~\bibnamefont
  {Guo}}, \bibinfo {author} {\bibfnamefont {A.~F.}\ \bibnamefont {Kockum}},
  \bibinfo {author} {\bibfnamefont {F.}~\bibnamefont {Marquardt}},\ and\
  \bibinfo {author} {\bibfnamefont {G.}~\bibnamefont {Johansson}},\ }\bibfield
  {title} {\bibinfo {title} {Oscillating bound states for a giant atom},\
  }\href {https://doi.org/10.1103/PhysRevResearch.2.043014} {\bibfield
  {journal} {\bibinfo  {journal} {Phys. Rev. Res.}\ }\textbf {\bibinfo {volume}
  {2}},\ \bibinfo {pages} {043014} (\bibinfo {year}
  {2020}{\natexlab{c}})}\BibitemShut {NoStop}%
\bibitem [{\citenamefont {Leggett}\ \emph {et~al.}(1987)\citenamefont
  {Leggett}, \citenamefont {Chakravarty}, \citenamefont {Dorsey}, \citenamefont
  {Fisher}, \citenamefont {Garg},\ and\ \citenamefont {Zwerger}}]{Leggett1987}%
  \BibitemOpen
  \bibfield  {author} {\bibinfo {author} {\bibfnamefont {A.~J.}\ \bibnamefont
  {Leggett}}, \bibinfo {author} {\bibfnamefont {S.}~\bibnamefont
  {Chakravarty}}, \bibinfo {author} {\bibfnamefont {A.~T.}\ \bibnamefont
  {Dorsey}}, \bibinfo {author} {\bibfnamefont {M.~P.~A.}\ \bibnamefont
  {Fisher}}, \bibinfo {author} {\bibfnamefont {A.}~\bibnamefont {Garg}},\ and\
  \bibinfo {author} {\bibfnamefont {W.}~\bibnamefont {Zwerger}},\ }\bibfield
  {title} {\bibinfo {title} {Dynamics of the dissipative two-state system},\
  }\href {https://doi.org/10.1103/RevModPhys.59.1} {\bibfield  {journal}
  {\bibinfo  {journal} {Rev. Mod. Phys.}\ }\textbf {\bibinfo {volume} {59}},\
  \bibinfo {pages} {1} (\bibinfo {year} {1987})}\BibitemShut {NoStop}%
\bibitem [{\citenamefont {Tanimura}(1990)}]{Tanimura1990}%
  \BibitemOpen
  \bibfield  {author} {\bibinfo {author} {\bibfnamefont {Y.}~\bibnamefont
  {Tanimura}},\ }\bibfield  {title} {\bibinfo {title} {Nonperturbative
  expansion method for a quantum system coupled to a harmonic-oscillator
  bath},\ }\href {https://doi.org/10.1103/PhysRevA.41.6676} {\bibfield
  {journal} {\bibinfo  {journal} {Phys. Rev. A}\ }\textbf {\bibinfo {volume}
  {41}},\ \bibinfo {pages} {6676} (\bibinfo {year} {1990})}\BibitemShut
  {NoStop}%
\bibitem [{\citenamefont {Ishizaki}\ and\ \citenamefont
  {Tanimura}(2005)}]{Ishizaki2005}%
  \BibitemOpen
  \bibfield  {author} {\bibinfo {author} {\bibfnamefont {A.}~\bibnamefont
  {Ishizaki}}\ and\ \bibinfo {author} {\bibfnamefont {Y.}~\bibnamefont
  {Tanimura}},\ }\bibfield  {title} {\bibinfo {title} {Quantum dynamics of
  system strongly coupled to low-temperature colored noise bath: Reduced
  hierarchy equations approach},\ }\href {https://doi.org/10.1143/JPSJ.74.3131}
  {\bibfield  {journal} {\bibinfo  {journal} {Journal of the Physical Society
  of Japan}\ }\textbf {\bibinfo {volume} {74}},\ \bibinfo {pages} {3131}
  (\bibinfo {year} {2005})}\BibitemShut {NoStop}%
\bibitem [{\citenamefont {Ding}\ \emph {et~al.}(2012)\citenamefont {Ding},
  \citenamefont {Xu},\ and\ \citenamefont {Yan}}]{Ding2012}%
  \BibitemOpen
  \bibfield  {author} {\bibinfo {author} {\bibfnamefont {J.-J.}\ \bibnamefont
  {Ding}}, \bibinfo {author} {\bibfnamefont {R.-X.}\ \bibnamefont {Xu}},\ and\
  \bibinfo {author} {\bibfnamefont {Y.}~\bibnamefont {Yan}},\ }\bibfield
  {title} {\bibinfo {title} {Optimizing hierarchical equations of motion for
  quantum dissipation and quantifying quantum bath effects on quantum transfer
  mechanisms},\ }\href {https://doi.org/10.1063/1.4724193} {\bibfield
  {journal} {\bibinfo  {journal} {The Journal of Chemical Physics}\ }\textbf
  {\bibinfo {volume} {136}},\ \bibinfo {pages} {224103} (\bibinfo {year}
  {2012})}\BibitemShut {NoStop}%
\bibitem [{\citenamefont {Hu}\ \emph {et~al.}(2010)\citenamefont {Hu},
  \citenamefont {Xu},\ and\ \citenamefont {Yan}}]{Hu2010}%
  \BibitemOpen
  \bibfield  {author} {\bibinfo {author} {\bibfnamefont {J.}~\bibnamefont
  {Hu}}, \bibinfo {author} {\bibfnamefont {R.-X.}\ \bibnamefont {Xu}},\ and\
  \bibinfo {author} {\bibfnamefont {Y.}~\bibnamefont {Yan}},\ }\bibfield
  {title} {\bibinfo {title} {Communication: Padé spectrum decomposition of
  fermi function and bose function},\ }\href
  {https://doi.org/10.1063/1.3484491} {\bibfield  {journal} {\bibinfo
  {journal} {The Journal of Chemical Physics}\ }\textbf {\bibinfo {volume}
  {133}},\ \bibinfo {pages} {101106} (\bibinfo {year} {2010})}\BibitemShut
  {NoStop}%
\bibitem [{\citenamefont {Hu}\ \emph {et~al.}(2011)\citenamefont {Hu},
  \citenamefont {Luo}, \citenamefont {Jiang}, \citenamefont {Xu},\ and\
  \citenamefont {Yan}}]{Hu2011}%
  \BibitemOpen
  \bibfield  {author} {\bibinfo {author} {\bibfnamefont {J.}~\bibnamefont
  {Hu}}, \bibinfo {author} {\bibfnamefont {M.}~\bibnamefont {Luo}}, \bibinfo
  {author} {\bibfnamefont {F.}~\bibnamefont {Jiang}}, \bibinfo {author}
  {\bibfnamefont {R.-X.}\ \bibnamefont {Xu}},\ and\ \bibinfo {author}
  {\bibfnamefont {Y.}~\bibnamefont {Yan}},\ }\bibfield  {title} {\bibinfo
  {title} {Padé spectrum decompositions of quantum distribution functions and
  optimal hierarchical equations of motion construction for quantum open
  systems},\ }\href {https://doi.org/10.1063/1.3602466} {\bibfield  {journal}
  {\bibinfo  {journal} {The Journal of Chemical Physics}\ }\textbf {\bibinfo
  {volume} {134}},\ \bibinfo {pages} {244106} (\bibinfo {year}
  {2011})}\BibitemShut {NoStop}%
\bibitem [{\citenamefont {Cui}\ \emph {et~al.}(2019)\citenamefont {Cui},
  \citenamefont {Zhang}, \citenamefont {Zheng}, \citenamefont {Xu},\ and\
  \citenamefont {Yan}}]{Cui2019}%
  \BibitemOpen
  \bibfield  {author} {\bibinfo {author} {\bibfnamefont {L.}~\bibnamefont
  {Cui}}, \bibinfo {author} {\bibfnamefont {H.-D.}\ \bibnamefont {Zhang}},
  \bibinfo {author} {\bibfnamefont {X.}~\bibnamefont {Zheng}}, \bibinfo
  {author} {\bibfnamefont {R.-X.}\ \bibnamefont {Xu}},\ and\ \bibinfo {author}
  {\bibfnamefont {Y.}~\bibnamefont {Yan}},\ }\bibfield  {title} {\bibinfo
  {title} {Highly efficient and accurate sum-over-poles expansion of fermi and
  bose functions at near zero temperatures: Fano spectrum decomposition
  scheme},\ }\href {https://doi.org/10.1063/1.5096945} {\bibfield  {journal}
  {\bibinfo  {journal} {The Journal of Chemical Physics}\ }\textbf {\bibinfo
  {volume} {151}},\ \bibinfo {pages} {024110} (\bibinfo {year}
  {2019})}\BibitemShut {NoStop}%
\bibitem [{\citenamefont {Zhang}\ \emph {et~al.}(2020)\citenamefont {Zhang},
  \citenamefont {Cui}, \citenamefont {Gong}, \citenamefont {Xu}, \citenamefont
  {Zheng},\ and\ \citenamefont {Yan}}]{Zhang2020}%
  \BibitemOpen
  \bibfield  {author} {\bibinfo {author} {\bibfnamefont {H.-D.}\ \bibnamefont
  {Zhang}}, \bibinfo {author} {\bibfnamefont {L.}~\bibnamefont {Cui}}, \bibinfo
  {author} {\bibfnamefont {H.}~\bibnamefont {Gong}}, \bibinfo {author}
  {\bibfnamefont {R.-X.}\ \bibnamefont {Xu}}, \bibinfo {author} {\bibfnamefont
  {X.}~\bibnamefont {Zheng}},\ and\ \bibinfo {author} {\bibfnamefont
  {Y.}~\bibnamefont {Yan}},\ }\bibfield  {title} {\bibinfo {title}
  {Hierarchical equations of motion method based on fano spectrum decomposition
  for low temperature environments},\ }\href
  {https://doi.org/10.1063/1.5136093} {\bibfield  {journal} {\bibinfo
  {journal} {The Journal of Chemical Physics}\ }\textbf {\bibinfo {volume}
  {152}},\ \bibinfo {pages} {064107} (\bibinfo {year} {2020})}\BibitemShut
  {NoStop}%
\bibitem [{\citenamefont {Xu}\ \emph {et~al.}(2022)\citenamefont {Xu},
  \citenamefont {Yan}, \citenamefont {Shi}, \citenamefont {Ankerhold},\ and\
  \citenamefont {Stockburger}}]{Xu2022}%
  \BibitemOpen
  \bibfield  {author} {\bibinfo {author} {\bibfnamefont {M.}~\bibnamefont
  {Xu}}, \bibinfo {author} {\bibfnamefont {Y.}~\bibnamefont {Yan}}, \bibinfo
  {author} {\bibfnamefont {Q.}~\bibnamefont {Shi}}, \bibinfo {author}
  {\bibfnamefont {J.}~\bibnamefont {Ankerhold}},\ and\ \bibinfo {author}
  {\bibfnamefont {J.~T.}\ \bibnamefont {Stockburger}},\ }\bibfield  {title}
  {\bibinfo {title} {Taming quantum noise for efficient low temperature
  simulations of open quantum systems},\ }\href
  {https://doi.org/10.1103/PhysRevLett.129.230601} {\bibfield  {journal}
  {\bibinfo  {journal} {Phys. Rev. Lett.}\ }\textbf {\bibinfo {volume} {129}},\
  \bibinfo {pages} {230601} (\bibinfo {year} {2022})}\BibitemShut {NoStop}%
\bibitem [{\citenamefont {Vanhuffel}\ \emph {et~al.}(1994)\citenamefont
  {Vanhuffel}, \citenamefont {Chen}, \citenamefont {Decanniere},\ and\
  \citenamefont {Vanhecke}}]{Vanhuffel1994}%
  \BibitemOpen
  \bibfield  {author} {\bibinfo {author} {\bibfnamefont {S.}~\bibnamefont
  {Vanhuffel}}, \bibinfo {author} {\bibfnamefont {H.}~\bibnamefont {Chen}},
  \bibinfo {author} {\bibfnamefont {C.}~\bibnamefont {Decanniere}},\ and\
  \bibinfo {author} {\bibfnamefont {P.}~\bibnamefont {Vanhecke}},\ }\bibfield
  {title} {\bibinfo {title} {Algorithm for time-domain nmr data fitting based
  on total least squares},\ }\href
  {https://doi.org/https://doi.org/10.1006/jmra.1994.1209} {\bibfield
  {journal} {\bibinfo  {journal} {Journal of Magnetic Resonance, Series A}\
  }\textbf {\bibinfo {volume} {110}},\ \bibinfo {pages} {228} (\bibinfo {year}
  {1994})}\BibitemShut {NoStop}%
\bibitem [{\citenamefont {Aushev}\ \emph {et~al.}(2014)\citenamefont {Aushev},
  \citenamefont {Kozhinov},\ and\ \citenamefont {Tsybin}}]{Aushev2014}%
  \BibitemOpen
  \bibfield  {author} {\bibinfo {author} {\bibfnamefont {T.}~\bibnamefont
  {Aushev}}, \bibinfo {author} {\bibfnamefont {A.~N.}\ \bibnamefont
  {Kozhinov}},\ and\ \bibinfo {author} {\bibfnamefont {Y.~O.}\ \bibnamefont
  {Tsybin}},\ }\bibfield  {title} {\bibinfo {title} {Least-squares fitting of
  time-domain signals for fourier transform mass spectrometry},\ }\href
  {https://doi.org/10.1007/s13361-014-0888-x} {\bibfield  {journal} {\bibinfo
  {journal} {Journal of the American Society for Mass Spectrometry}\ }\textbf
  {\bibinfo {volume} {25}},\ \bibinfo {pages} {1263} (\bibinfo {year}
  {2014})},\ \bibinfo {note} {pMID: 24789745}\BibitemShut {NoStop}%
\bibitem [{\citenamefont {Hartmann}\ \emph {et~al.}(2019)\citenamefont
  {Hartmann}, \citenamefont {Werther}, \citenamefont {Grossmann},\ and\
  \citenamefont {Strunz}}]{Hartmann2019}%
  \BibitemOpen
  \bibfield  {author} {\bibinfo {author} {\bibfnamefont {R.}~\bibnamefont
  {Hartmann}}, \bibinfo {author} {\bibfnamefont {M.}~\bibnamefont {Werther}},
  \bibinfo {author} {\bibfnamefont {F.}~\bibnamefont {Grossmann}},\ and\
  \bibinfo {author} {\bibfnamefont {W.~T.}\ \bibnamefont {Strunz}},\ }\bibfield
   {title} {\bibinfo {title} {Exact open quantum system dynamics: Optimal
  frequency vs time representation of bath correlations},\ }\href
  {https://doi.org/10.1063/1.5097158} {\bibfield  {journal} {\bibinfo
  {journal} {The Journal of Chemical Physics}\ }\textbf {\bibinfo {volume}
  {150}},\ \bibinfo {pages} {234105} (\bibinfo {year} {2019})}\BibitemShut
  {NoStop}%
\bibitem [{\citenamefont {Chen}\ \emph {et~al.}(2022)\citenamefont {Chen},
  \citenamefont {Wang}, \citenamefont {Zheng}, \citenamefont {Xu},\ and\
  \citenamefont {Yan}}]{Chen2022}%
  \BibitemOpen
  \bibfield  {author} {\bibinfo {author} {\bibfnamefont {Z.-H.}\ \bibnamefont
  {Chen}}, \bibinfo {author} {\bibfnamefont {Y.}~\bibnamefont {Wang}}, \bibinfo
  {author} {\bibfnamefont {X.}~\bibnamefont {Zheng}}, \bibinfo {author}
  {\bibfnamefont {R.-X.}\ \bibnamefont {Xu}},\ and\ \bibinfo {author}
  {\bibfnamefont {Y.}~\bibnamefont {Yan}},\ }\bibfield  {title} {\bibinfo
  {title} {Universal time-domain prony fitting decomposition for optimized
  hierarchical quantum master equations},\ }\href
  {https://doi.org/10.1063/5.0095961} {\bibfield  {journal} {\bibinfo
  {journal} {The Journal of Chemical Physics}\ }\textbf {\bibinfo {volume}
  {156}},\ \bibinfo {pages} {221102} (\bibinfo {year} {2022})}\BibitemShut
  {NoStop}%
\bibitem [{\citenamefont {Schaubert}(1979)}]{Schaubert1979}%
  \BibitemOpen
  \bibfield  {author} {\bibinfo {author} {\bibfnamefont {D.}~\bibnamefont
  {Schaubert}},\ }\bibfield  {title} {\bibinfo {title} {Application of prony's
  method to time-domain reflectometer data and equivalent circuit synthesis},\
  }\href {https://doi.org/10.1109/TAP.1979.1142060} {\bibfield  {journal}
  {\bibinfo  {journal} {IEEE Transactions on Antennas and Propagation}\
  }\textbf {\bibinfo {volume} {27}},\ \bibinfo {pages} {180} (\bibinfo {year}
  {1979})}\BibitemShut {NoStop}%
\bibitem [{\citenamefont {Abramowitz}\ and\ \citenamefont
  {Stegun}(1965)}]{Abramowitz1965}%
  \BibitemOpen
  \bibfield  {author} {\bibinfo {author} {\bibfnamefont {M.}~\bibnamefont
  {Abramowitz}}\ and\ \bibinfo {author} {\bibfnamefont {I.}~\bibnamefont
  {Stegun}},\ }\href {https://books.google.de/books?id=MtU8uP7XMvoC} {\emph
  {\bibinfo {title} {Handbook of Mathematical Functions: With Formulas, Graphs,
  and Mathematical Tables}}},\ Applied mathematics series\ (\bibinfo
  {publisher} {Dover Publications},\ \bibinfo {year} {1965})\BibitemShut
  {NoStop}%
\bibitem [{\citenamefont {Van~Kampen}(2011)}]{vanKampen2011}%
  \BibitemOpen
  \bibfield  {author} {\bibinfo {author} {\bibfnamefont {N.}~\bibnamefont
  {Van~Kampen}},\ }\href {https://books.google.de/books?id=N6II-6HlPxEC} {\emph
  {\bibinfo {title} {Stochastic Processes in Physics and Chemistry}}},\
  North-Holland Personal Library\ (\bibinfo  {publisher} {North Holland},\
  \bibinfo {year} {2011})\BibitemShut {NoStop}%
\bibitem [{\citenamefont {Hall}\ \emph {et~al.}(2014)\citenamefont {Hall},
  \citenamefont {Cresser}, \citenamefont {Li},\ and\ \citenamefont
  {Andersson}}]{Hall2014}%
  \BibitemOpen
  \bibfield  {author} {\bibinfo {author} {\bibfnamefont {M.~J.~W.}\
  \bibnamefont {Hall}}, \bibinfo {author} {\bibfnamefont {J.~D.}\ \bibnamefont
  {Cresser}}, \bibinfo {author} {\bibfnamefont {L.}~\bibnamefont {Li}},\ and\
  \bibinfo {author} {\bibfnamefont {E.}~\bibnamefont {Andersson}},\ }\bibfield
  {title} {\bibinfo {title} {Canonical form of master equations and
  characterization of non-markovianity},\ }\href
  {https://doi.org/10.1103/PhysRevA.89.042120} {\bibfield  {journal} {\bibinfo
  {journal} {Phys. Rev. A}\ }\textbf {\bibinfo {volume} {89}},\ \bibinfo
  {pages} {042120} (\bibinfo {year} {2014})}\BibitemShut {NoStop}%
\bibitem [{\citenamefont {Hartmann}\ and\ \citenamefont
  {Strunz}(2020)}]{Hartmann2020}%
  \BibitemOpen
  \bibfield  {author} {\bibinfo {author} {\bibfnamefont {R.}~\bibnamefont
  {Hartmann}}\ and\ \bibinfo {author} {\bibfnamefont {W.~T.}\ \bibnamefont
  {Strunz}},\ }\bibfield  {title} {\bibinfo {title} {Accuracy assessment of
  perturbative master equations: Embracing nonpositivity},\ }\href
  {https://doi.org/10.1103/PhysRevA.101.012103} {\bibfield  {journal} {\bibinfo
   {journal} {Phys. Rev. A}\ }\textbf {\bibinfo {volume} {101}},\ \bibinfo
  {pages} {012103} (\bibinfo {year} {2020})}\BibitemShut {NoStop}%
\bibitem [{\citenamefont {Scali}\ \emph {et~al.}(2021)\citenamefont {Scali},
  \citenamefont {Anders},\ and\ \citenamefont {Correa}}]{Scali2021}%
  \BibitemOpen
  \bibfield  {author} {\bibinfo {author} {\bibfnamefont {S.}~\bibnamefont
  {Scali}}, \bibinfo {author} {\bibfnamefont {J.}~\bibnamefont {Anders}},\ and\
  \bibinfo {author} {\bibfnamefont {L.~A.}\ \bibnamefont {Correa}},\ }\bibfield
   {title} {\bibinfo {title} {Local master equations bypass the secular
  approximation},\ }\href {https://doi.org/10.22331/q-2021-05-01-451}
  {\bibfield  {journal} {\bibinfo  {journal} {{Quantum}}\ }\textbf {\bibinfo
  {volume} {5}},\ \bibinfo {pages} {451} (\bibinfo {year} {2021})}\BibitemShut
  {NoStop}%
\bibitem [{\citenamefont {Frisk~Kockum}(2021)}]{Anton2021}%
  \BibitemOpen
  \bibfield  {author} {\bibinfo {author} {\bibfnamefont {A.}~\bibnamefont
  {Frisk~Kockum}},\ }\bibfield  {title} {\bibinfo {title} {Quantum optics with
  giant atoms---the first five years},\ }in\ \href@noop {} {\emph {\bibinfo
  {booktitle} {International Symposium on Mathematics, Quantum Theory, and
  Cryptography}}},\ \bibinfo {editor} {edited by\ \bibinfo {editor}
  {\bibfnamefont {T.}~\bibnamefont {Takagi}}, \bibinfo {editor} {\bibfnamefont
  {M.}~\bibnamefont {Wakayama}}, \bibinfo {editor} {\bibfnamefont
  {K.}~\bibnamefont {Tanaka}}, \bibinfo {editor} {\bibfnamefont
  {N.}~\bibnamefont {Kunihiro}}, \bibinfo {editor} {\bibfnamefont
  {K.}~\bibnamefont {Kimoto}},\ and\ \bibinfo {editor} {\bibfnamefont
  {Y.}~\bibnamefont {Ikematsu}}}\ (\bibinfo  {publisher} {Springer Singapore},\
  \bibinfo {address} {Singapore},\ \bibinfo {year} {2021})\ pp.\ \bibinfo
  {pages} {125--146}\BibitemShut {NoStop}%
\bibitem [{\citenamefont {Chu}\ \emph {et~al.}(2017)\citenamefont {Chu},
  \citenamefont {Kharel}, \citenamefont {Renninger}, \citenamefont {Burkhart},
  \citenamefont {Frunzio}, \citenamefont {Rakich},\ and\ \citenamefont
  {Schoelkopf}}]{Chu2017}%
  \BibitemOpen
  \bibfield  {author} {\bibinfo {author} {\bibfnamefont {Y.}~\bibnamefont
  {Chu}}, \bibinfo {author} {\bibfnamefont {P.}~\bibnamefont {Kharel}},
  \bibinfo {author} {\bibfnamefont {W.~H.}\ \bibnamefont {Renninger}}, \bibinfo
  {author} {\bibfnamefont {L.~D.}\ \bibnamefont {Burkhart}}, \bibinfo {author}
  {\bibfnamefont {L.}~\bibnamefont {Frunzio}}, \bibinfo {author} {\bibfnamefont
  {P.~T.}\ \bibnamefont {Rakich}},\ and\ \bibinfo {author} {\bibfnamefont
  {R.~J.}\ \bibnamefont {Schoelkopf}},\ }\bibfield  {title} {\bibinfo {title}
  {Quantum acoustics with superconducting qubits},\ }\href
  {https://doi.org/10.1126/science.aao1511} {\bibfield  {journal} {\bibinfo
  {journal} {Science}\ }\textbf {\bibinfo {volume} {358}},\ \bibinfo {pages}
  {199} (\bibinfo {year} {2017})}\BibitemShut {NoStop}%
\bibitem [{\citenamefont {Guarnieri}\ \emph {et~al.}(2018)\citenamefont
  {Guarnieri}, \citenamefont {Kol\'a\ifmmode~\check{r}\else \v{r}\fi{}},\ and\
  \citenamefont {Filip}}]{Guarnieri2018}%
  \BibitemOpen
  \bibfield  {author} {\bibinfo {author} {\bibfnamefont {G.}~\bibnamefont
  {Guarnieri}}, \bibinfo {author} {\bibfnamefont {M.}~\bibnamefont
  {Kol\'a\ifmmode~\check{r}\else \v{r}\fi{}}},\ and\ \bibinfo {author}
  {\bibfnamefont {R.}~\bibnamefont {Filip}},\ }\bibfield  {title} {\bibinfo
  {title} {Steady-state coherences by composite system-bath interactions},\
  }\href {https://doi.org/10.1103/PhysRevLett.121.070401} {\bibfield  {journal}
  {\bibinfo  {journal} {Phys. Rev. Lett.}\ }\textbf {\bibinfo {volume} {121}},\
  \bibinfo {pages} {070401} (\bibinfo {year} {2018})}\BibitemShut {NoStop}%
\bibitem [{\citenamefont {Purkayastha}\ \emph {et~al.}(2020)\citenamefont
  {Purkayastha}, \citenamefont {Guarnieri}, \citenamefont {Mitchison},
  \citenamefont {Filip},\ and\ \citenamefont {Goold}}]{Purkayastha2020}%
  \BibitemOpen
  \bibfield  {author} {\bibinfo {author} {\bibfnamefont {A.}~\bibnamefont
  {Purkayastha}}, \bibinfo {author} {\bibfnamefont {G.}~\bibnamefont
  {Guarnieri}}, \bibinfo {author} {\bibfnamefont {M.~T.}\ \bibnamefont
  {Mitchison}}, \bibinfo {author} {\bibfnamefont {R.}~\bibnamefont {Filip}},\
  and\ \bibinfo {author} {\bibfnamefont {J.}~\bibnamefont {Goold}},\ }\bibfield
   {title} {\bibinfo {title} {Tunable phonon-induced steady-state coherence in
  a double-quantum-dot charge qubit},\ }\href
  {https://doi.org/https://doi.org/10.1038/s41534-020-0256-6} {\bibfield
  {journal} {\bibinfo  {journal} {npj Quantum Information}\ }\textbf {\bibinfo
  {volume} {6}},\ \bibinfo {pages} {27} (\bibinfo {year} {2020})}\BibitemShut
  {NoStop}%
\bibitem [{\citenamefont {Campisi}\ \emph {et~al.}(2009)\citenamefont
  {Campisi}, \citenamefont {Talkner},\ and\ \citenamefont
  {H\"anggi}}]{Hanggi2009}%
  \BibitemOpen
  \bibfield  {author} {\bibinfo {author} {\bibfnamefont {M.}~\bibnamefont
  {Campisi}}, \bibinfo {author} {\bibfnamefont {P.}~\bibnamefont {Talkner}},\
  and\ \bibinfo {author} {\bibfnamefont {P.}~\bibnamefont {H\"anggi}},\
  }\bibfield  {title} {\bibinfo {title} {Fluctuation theorem for arbitrary open
  quantum systems},\ }\href {https://doi.org/10.1103/PhysRevLett.102.210401}
  {\bibfield  {journal} {\bibinfo  {journal} {Phys. Rev. Lett.}\ }\textbf
  {\bibinfo {volume} {102}},\ \bibinfo {pages} {210401} (\bibinfo {year}
  {2009})}\BibitemShut {NoStop}%
\bibitem [{\citenamefont {Rivas}(2020)}]{Rivas2020}%
  \BibitemOpen
  \bibfield  {author} {\bibinfo {author} {\bibfnamefont {A.}~\bibnamefont
  {Rivas}},\ }\bibfield  {title} {\bibinfo {title} {Strong coupling
  thermodynamics of open quantum systems},\ }\href
  {https://doi.org/10.1103/PhysRevLett.124.160601} {\bibfield  {journal}
  {\bibinfo  {journal} {Phys. Rev. Lett.}\ }\textbf {\bibinfo {volume} {124}},\
  \bibinfo {pages} {160601} (\bibinfo {year} {2020})}\BibitemShut {NoStop}%
\bibitem [{\citenamefont {Li}\ and\ \citenamefont {Cao}(2024)}]{Li2024}%
  \BibitemOpen
  \bibfield  {author} {\bibinfo {author} {\bibfnamefont {L.}~\bibnamefont
  {Li}}\ and\ \bibinfo {author} {\bibfnamefont {X.}~\bibnamefont {Cao}},\
  }\bibfield  {title} {\bibinfo {title} {Coherent destruction of tunneling of
  quantum energy transport in a driven nonequilibrium spin-boson model},\
  }\href {https://doi.org/10.1103/PhysRevB.110.075403} {\bibfield  {journal}
  {\bibinfo  {journal} {Phys. Rev. B}\ }\textbf {\bibinfo {volume} {110}},\
  \bibinfo {pages} {075403} (\bibinfo {year} {2024})}\BibitemShut {NoStop}%
\bibitem [{\citenamefont {Ritter}\ \emph {et~al.}(2025)\citenamefont {Ritter},
  \citenamefont {Long}, \citenamefont {Yue}, \citenamefont {Chandran},\ and\
  \citenamefont {Koll\'ar}}]{Ritter2025}%
  \BibitemOpen
  \bibfield  {author} {\bibinfo {author} {\bibfnamefont {M.}~\bibnamefont
  {Ritter}}, \bibinfo {author} {\bibfnamefont {D.~M.}\ \bibnamefont {Long}},
  \bibinfo {author} {\bibfnamefont {Q.}~\bibnamefont {Yue}}, \bibinfo {author}
  {\bibfnamefont {A.}~\bibnamefont {Chandran}},\ and\ \bibinfo {author}
  {\bibfnamefont {A.~J.}\ \bibnamefont {Koll\'ar}},\ }\bibfield  {title}
  {\bibinfo {title} {Autonomous stabilization of floquet states using static
  dissipation},\ }\href {https://doi.org/10.1103/3dpp-p2pr} {\bibfield
  {journal} {\bibinfo  {journal} {Phys. Rev. X}\ }\textbf {\bibinfo {volume}
  {15}},\ \bibinfo {pages} {031028} (\bibinfo {year} {2025})}\BibitemShut
  {NoStop}%
\bibitem [{\citenamefont {B\"acker}\ \emph {et~al.}(2024)\citenamefont
  {B\"acker}, \citenamefont {Beyer},\ and\ \citenamefont
  {Strunz}}]{Backer2024}%
  \BibitemOpen
  \bibfield  {author} {\bibinfo {author} {\bibfnamefont {C.}~\bibnamefont
  {B\"acker}}, \bibinfo {author} {\bibfnamefont {K.}~\bibnamefont {Beyer}},\
  and\ \bibinfo {author} {\bibfnamefont {W.~T.}\ \bibnamefont {Strunz}},\
  }\bibfield  {title} {\bibinfo {title} {Local disclosure of quantum memory in
  non-markovian dynamics},\ }\href
  {https://doi.org/10.1103/PhysRevLett.132.060402} {\bibfield  {journal}
  {\bibinfo  {journal} {Phys. Rev. Lett.}\ }\textbf {\bibinfo {volume} {132}},\
  \bibinfo {pages} {060402} (\bibinfo {year} {2024})}\BibitemShut {NoStop}%
\bibitem [{\citenamefont {Breuer}\ and\ \citenamefont
  {Petruccione}(2002)}]{Breuer2002}%
  \BibitemOpen
  \bibfield  {author} {\bibinfo {author} {\bibfnamefont {H.}~\bibnamefont
  {Breuer}}\ and\ \bibinfo {author} {\bibfnamefont {F.}~\bibnamefont
  {Petruccione}},\ }\href {https://books.google.de/books?id=0Yx5VzaMYm8C}
  {\emph {\bibinfo {title} {The Theory of Open Quantum Systems}}}\ (\bibinfo
  {publisher} {Oxford University Press},\ \bibinfo {year} {2002})\BibitemShut
  {NoStop}%
\bibitem [{\citenamefont {Cheng}\ and\ \citenamefont
  {Fleming}(2009)}]{Cheng2009}%
  \BibitemOpen
  \bibfield  {author} {\bibinfo {author} {\bibfnamefont {Y.-C.}\ \bibnamefont
  {Cheng}}\ and\ \bibinfo {author} {\bibfnamefont {G.~R.}\ \bibnamefont
  {Fleming}},\ }\bibfield  {title} {\bibinfo {title} {Dynamics of light
  harvesting in photosynthesis},\ }\href
  {https://doi.org/https://doi.org/10.1146/annurev.physchem.040808.090259}
  {\bibfield  {journal} {\bibinfo  {journal} {Annual Review of Physical
  Chemistry}\ }\textbf {\bibinfo {volume} {60}},\ \bibinfo {pages} {241}
  (\bibinfo {year} {2009})}\BibitemShut {NoStop}%
\bibitem [{\citenamefont {Ritschel}\ \emph {et~al.}(2011)\citenamefont
  {Ritschel}, \citenamefont {Roden}, \citenamefont {Strunz},\ and\
  \citenamefont {Eisfeld}}]{Ritschel2011}%
  \BibitemOpen
  \bibfield  {author} {\bibinfo {author} {\bibfnamefont {G.}~\bibnamefont
  {Ritschel}}, \bibinfo {author} {\bibfnamefont {J.}~\bibnamefont {Roden}},
  \bibinfo {author} {\bibfnamefont {W.~T.}\ \bibnamefont {Strunz}},\ and\
  \bibinfo {author} {\bibfnamefont {A.}~\bibnamefont {Eisfeld}},\ }\bibfield
  {title} {\bibinfo {title} {An efficient method to calculate excitation energy
  transfer in light-harvesting systems: application to the
  fenna–matthews–olson complex},\ }\href
  {https://doi.org/10.1088/1367-2630/13/11/113034} {\bibfield  {journal}
  {\bibinfo  {journal} {New Journal of Physics}\ }\textbf {\bibinfo {volume}
  {13}},\ \bibinfo {pages} {113034} (\bibinfo {year} {2011})}\BibitemShut
  {NoStop}%
\bibitem [{\citenamefont {\L{}obejko}\ \emph {et~al.}(2024)\citenamefont
  {\L{}obejko}, \citenamefont {Winczewski}, \citenamefont {Su\'arez},
  \citenamefont {Alicki},\ and\ \citenamefont {Horodecki}}]{Lobejko2024}%
  \BibitemOpen
  \bibfield  {author} {\bibinfo {author} {\bibfnamefont {M.}~\bibnamefont
  {\L{}obejko}}, \bibinfo {author} {\bibfnamefont {M.}~\bibnamefont
  {Winczewski}}, \bibinfo {author} {\bibfnamefont {G.}~\bibnamefont
  {Su\'arez}}, \bibinfo {author} {\bibfnamefont {R.}~\bibnamefont {Alicki}},\
  and\ \bibinfo {author} {\bibfnamefont {M.}~\bibnamefont {Horodecki}},\
  }\bibfield  {title} {\bibinfo {title} {Corrections to the hamiltonian induced
  by finite-strength coupling to the environment},\ }\href
  {https://doi.org/10.1103/PhysRevE.110.014144} {\bibfield  {journal} {\bibinfo
   {journal} {Phys. Rev. E}\ }\textbf {\bibinfo {volume} {110}},\ \bibinfo
  {pages} {014144} (\bibinfo {year} {2024})}\BibitemShut {NoStop}%
\end{thebibliography}
\end{document}